\documentclass[prl,aps,twocolumn,preprintnumbers,amssymb,nobibnotes,nofootinbib,raggedbottom,10pt]{revtex4-1}
\usepackage{amsmath}
\usepackage{hyperref,graphicx,array,multirow,color,amsmath,mathrsfs,titlesec,shuffle,tikz}
\usepackage{float}
\usepackage{mathtools}
\usepackage{xcolor}
\usepackage{cancel}
\usepackage{tikzit}

\tikzstyle{med 2}=[fill={rgb,255: red,233; green,233; blue,233}, draw=black, shape=circle, tikzit shape=circle, minimum size=1cm, thick]
\tikzstyle{med dark}=[fill={rgb,255: red,193; green,193; blue,193}, draw=black, shape=circle, tikzit shape=circle, minimum size=1.0cm, thick]
\tikzstyle{dot}=[fill=black, draw=black, shape=circle, tikzit shape=circle, inner sep=0.75pt]
\tikzstyle{medium blob}=[fill={rgb,255: red,237; green,237; blue,237}, draw=black, shape=circle, tikzit shape=circle, minimum size=1.07cm, thick]
\tikzstyle{NMHV}=[fill=white, draw=black, shape=circle, tikzit shape=circle, minimum size=0.3cm, thick]
\tikzstyle{MHV}=[fill=black, draw=black, shape=circle, tikzit shape=circle, minimum size=0.3cm, thick]
\tikzstyle{Large circle 2}=[fill=none, draw=black, shape=circle, tikzit shape=circle, minimum size=3.0cm, thick]
\tikzstyle{Large red}=[fill=none, draw={rgb,255: red,218; green,9; blue,9}, shape=circle, tikzit shape=circle, minimum size=2.4cm, thick, dashed]
\tikzstyle{medium red}=[fill=none, draw={rgb,255: red,218; green,9; blue,9}, shape=circle, tikzit shape=circle, minimum size=2cm, thick, dashed]
\tikzstyle{med black}=[fill=none, draw=black, shape=circle, tikzit shape=circle, thick, minimum size=2.5cm]
\tikzstyle{Small MHV}=[fill=black, draw=black, shape=circle, tikzit shape=circle, inner sep=0.07cm, thick]
\tikzstyle{Small NMHV}=[fill=white, draw=black, shape=circle, tikzit shape=circle, inner sep=0.1cm, thick]
\tikzstyle{Large circle}=[fill=none, draw={rgb,255: red,223; green,223; blue,223}, shape=circle, tikzit shape=circle, minimum size=3.7cm, line width=4.5mm]
\tikzstyle{Large Circle 3}=[fill=none, draw=black, shape=circle, minimum size=4.15cm, thick, tikzit shape=circle]
\tikzstyle{Large Circle 4}=[fill=none, draw=black, shape=circle, thick, minimum size=3.65cm]
\tikzstyle{Large circle 5}=[fill=none, draw=black, shape=circle, thick, minimum size=2.75cm]
\tikzstyle{Large circle 6}=[fill=none, draw=black, shape=circle, draw={rgb,255: red,223; green,223; blue,223}, minimum size=3.2cm, line width=4.5mm]
\tikzstyle{small circle}=[fill=white, draw=black, shape=circle, tikzit shape=circle, minimum size=1.0cm, thick]
\tikzstyle{sc 2}=[fill=white, draw=black, shape=circle, inner sep=0.06cm, thick, tikzit shape=circle]
\tikzstyle{new style 0}=[fill=none, draw={rgb,255: red,218; green,9; blue,9}, shape=circle, thick, dashed, tikzit draw={rgb,255: red,218; green,9; blue,9}, tikzit shape=circle, minimum size=0.2cm]
\tikzstyle{Cross}=[fill=white, draw=black, shape=circle, cross out, inner sep=0.06cm, thick]

\tikzstyle{Line}=[-, thick, fill={rgb,255: red,167; green,173; blue,221}, tikzit fill={rgb,255: red,198; green,198; blue,198}]
\tikzstyle{Arrow}=[fill=none, ->, -stealth, thick]
\tikzstyle{Red Dashed}=[-, dashed, draw={rgb,255: red,202; green,0; blue,0}, thick, fill=none]
\tikzstyle{RedArrow}=[fill=none, ->, -stealth, thick, draw={rgb,255: red,218; green,9; blue,9}]
\tikzstyle{Red Line}=[-, thick, draw={rgb,255: red,218; green,9; blue,9}]
\tikzstyle{Filled line}=[-, thick, fill=white]
\tikzstyle{white line}=[-, draw=white, line width=1.6mm]
\tikzstyle{Thick Red}=[-, draw={rgb,255: red,218; green,9; blue,9}, line width=0.6mm]
\tikzstyle{Blue dashed}=[-, draw={rgb,255: red,0; green,0; blue,166}, thick, dashed, fill=none]
\tikzstyle{Double arrow}=[<->, stealth-stealth, thick]
\tikzstyle{Blue}=[-, thick, draw={rgb,255: red,0; green,0; blue,166}]
\tikzstyle{Blue arrow}=[->, -stealth, thick, draw={rgb,255: red,0; green,0; blue,166}]
\tikzstyle{densely dotted}=[-, densely dashed, draw={rgb,255: red,218; green,9; blue,9}, thick]
\tikzstyle{Green arrow}=[->, -stealth, thick, draw={rgb,255: red,9; green,138; blue,0}]
\tikzstyle{Green dashed}=[-, thick, dashed, draw={rgb,255: red,9; green,138; blue,0}]

\newcommand{\Ddots}{\hbox to 1em{.\hss.\hss.\hss}}

\begin{document}

\title{Hidden Amplitude Zeros From Double Copy}

\author{Christoph Bartsch$^\Diamond$}
\author{Taro V. Brown$^\spadesuit$}
\author{Karol Kampf$\,^\Diamond$}
\author{Umut Oktem$^\spadesuit$}
\author{Shruti Paranjape$^\spadesuit$}
\author{Jaroslav Trnka$^{\spadesuit}$}

\affiliation{$^\spadesuit$Center for Quantum Mathematics and Physics (QMAP), University of California, Davis, CA, USA}
\affiliation{$^\Diamond$Institute for Particle and Nuclear Physics, Charles University, Prague, Czech Republic}

\begin{abstract} 
Recently, Arkani-Hamed et al. proposed the existence of zeros in scattering amplitudes in certain quantum field theories including the cubic adjoint scalar theory Tr($\phi^3$), the $SU(N)$ non-linear sigma model (NLSM) and Yang-Mills (YM) theory. These hidden zeros are special kinematic points where the amplitude vanishes and factorizes into a product of lower-point amplitudes, similar to factorization near poles. In this letter, we show a close connection between the existence of such zeros and color-kinematics duality. In fact, all zeros can be derived from the Bern-Carrasco-Johansson (BCJ) relations. We also show that these zeros extend via the Kawai-Lewellen-Tye (KLT) relations to special Galileon amplitudes and their corrections, evincing that these hidden zeros are also present in permutation-invariant amplitudes.
\end{abstract}

\maketitle

\vspace{-0.3cm}
\section{Introduction}

Scattering amplitudes in quantum field theory are constrained by physical principles. Tree-level amplitudes have poles which are well-understood from the local nature of interactions. Unitarity dictates that amplitudes factorize on these poles into products of lower-point amplitudes and this can be used to reconstruct the amplitude recursively \cite{Britto:2005fq,Cohen:2010mi,Cheung:2015cba}. In specific theories there are even more kinematic properties that scattering amplitudes can have, one very interesting set is the presence of \emph{zeros}. The most famous example is the \emph{Adler zero} \cite{Adler:1964um,Kampf:2013vha}, i.e. vanishing of the amplitude in the soft limit.
%
%
This is a consequence of the shift symmetry of the underlying Lagrangian, present in theories with a spontaneously broken symmetry. 
In fact, just factorization properties and the Adler zero are enough to completely fix all $n$-point amplitudes in the NLSM theory. Motivated by that, one can impose the behavior in the soft limit as a kinematic constraint and search for new (or already known) theories. Dubbed the \emph{soft bootstrap} \cite{Cheung:2014dqa,Cheung:2015ota,Cheung:2016drk,Elvang:2018dco}, this has led to many interesting insights and singled out certain special theories such as the NLSM, Born-Infeld and Galileon models as unique theories with specified soft behavior. These theories also play a special role in different contexts such as in the Cachazo-He-Yuan (CHY) scattering equations, ambitwistor strings or color-kinematics duality \cite{Cachazo:2014xea,Casali:2015vta,Bern:2008qj} and were also classified using algebraic methods \cite{Bogers:2018zeg,Roest:2019oiw}. One particular constraint in the bootstrap is an \emph{enhanced soft limit} \cite{Cheung:2014dqa} i.e. scaling $p_i\rightarrow tp_i$ and $t\rightarrow0$, requires
\begin{equation}
    \lim_{t\rightarrow 0} \mathcal{A}_n = {\cal O}(t^\sigma)\,. \label{enhanced}
\end{equation}
In multi-field theories, we discuss \emph{ordered} amplitudes $A_n$, where the full amplitude is decomposed into cyclic sectors with fixed orderings of external legs. For NLSM amplitudes such a decomposition is known as \emph{flavor ordering},
\begin{equation}
{\cal A}_n = \sum_{a\in S_{n}/Z_n} {\rm Tr}\,(T^{a_1}\dots T^{a_n})\,A_n[a_1a_2\dots a_n]\,,
\end{equation}
and the Adler zero ($\sigma=1$ soft limit) is then enjoyed by the full amplitude ${\cal A}_n$ as well as the ordered amplitudes $A_n$ \cite{Kampf:2012fn,Kampf:2013vha}. In theories without internal group structure, there are \emph{unordered} amplitudes. One notable example is the special Galileon, discovered from the bootstrap method as the unique theory with $\sigma=3$ soft behavior \cite{Cheung:2014dqa}, later explained by the presence of an enhanced symmetry of the Lagrangian \cite{Hinterbichler:2015pqa}. The same amplitude also naturally appears in the CHY formalism \cite{Cachazo:2014xea}. Using the double copy relations, two copies of NLSM produce the special Galileon amplitudes \cite{Cachazo:2014xea}. 
%
%

In recent remarkable work \cite{Arkani-Hamed:2023swr,Arkani-Hamed:2024nhp,Arkani-Hamed:2024yvu}, Arkani-Hamed, Cao, Dong, Figueiredo and He discovered that amplitudes in certain theories exhibit very different types of zeros: they vanish when certain Mandelstam invariants $s_{ij}=(p_i+p_j)^2$ go to zero. A similar zero has also been noted to exist in vector boson pair production \cite{Dixon:1999di}. More generally, these \emph{hidden zeros} originate from the kinematic mesh picture \cite{Arkani-Hamed:2019vag}, and exist in amplitudes of ${\rm Tr}(\phi^3)$ theory and the NLSM, but also to YM theory (where the zeros involve polarization vectors). At low multiplicity, the four-point amplitude vanishes if $s_{13}=0$, and the five-point if $s_{13}=s_{14}=0$. The six-point amplitude has two types of zeros, we refer to them as \emph{skinny} and \emph{rectangle}:
\begin{align}
\begin{split}
    C_1&=\lbrace s_{13}=s_{14}=s_{15}=0\rbrace,\\
    C_2&=\lbrace s_{14}=s_{15}=s_{24}=s_{25}=0 \rbrace,
    \label{6pC}
\end{split}
\end{align}
and similarly for all cyclic permutations of labels. 
More generally, at $n$ points the cyclically inequivalent zeros are
\begin{align}
    C_m=\begin{Bmatrix} 
    \hspace{-0.0cm}s_{1m+2}\, =\,s_{1m+3}\,=\ldots = s_{1n-1}\,\, = 0\\
    \hspace{-0.0cm}s_{2m+2}\, =\,s_{2m+3}\,=\ldots = s_{2n-1}\,\, = 0\\
    \hspace{-0.55cm}\vdots \hspace{1.4cm} \vdots \hspace{2.2cm} \vdots\\
    \hspace{-0.3cm}\hspace{0.3cm}s_{mm+2}=s_{mm+3}=\ldots = s_{mn-1} = 0
    \end{Bmatrix},
    \label{eq:constrCm}
\end{align}
where $m=1,\dots, \lfloor\frac{n}{2}\rfloor -1$.
Furthermore, the amplitudes factorize in the neighborhood of these zeros into a four-point prefactor (not necessarily an amplitude in the same theory) and a ``top'' and ``bottom'' amplitude,
\begin{align}
    A_n\big\rvert_{C_m} \xrightarrow{s^\ast \neq 0} A_4 \times A_{m+2,B} \times A_{n-m,T}\, ,
    \label{Afact}
\end{align}
as can also be read off directly from the kinematic mesh picture, with $s_{i\cdots n}\equiv (p_i+\cdots +p_n)^2$,
\begin{equation}
  A_n\big\rvert_{C_m} \xrightarrow{s^\ast \neq 0} \hspace{0.1cm}\begin{tikzpicture}
	\begin{pgfonlayer}{nodelayer}
		\node [style=none] (1) at (0, 4) {};
		\node [style=none] (2) at (4, 0) {};
		\node [style=none] (3) at (0, -4) {};
		\node [style=none] (4) at (2.5, 1.5) {};
		\node [style=none] (5) at (1.5, -2.5) {};
		\node [style=none] (6) at (0, -1) {};
		\node [style=none] (7) at (1.25, -0.5) {};
		\node [style=none] (8) at (2, 0.25) {};
		\node [style=none] (9) at (2.75, -0.5) {};
		\node [style=none] (10) at (2, -1.25) {};
		\node [style=none] (11) at (2, -0.5) {$s^*$};
		\node [style=none] (15) at (2.75, -3) {$s_{1\cdots m{+}1}$};
		\node [style=dot] (16) at (1.5, -2.5) {};
		\node [style=dot] (17) at (2.5, 1.5) {};
		\node [style=none] (18) at (4.2, 1.8) {$s_{m{+}1\cdots n{-}1}$};
		\node [style=none] (19) at (0.75, 1.5) {$\,\,A_T$};
		\node [style=none] (20) at (0.5, -2.5) {$\,\,A_B$};
	\end{pgfonlayer}
	\begin{pgfonlayer}{edgelayer}
		\draw [style=Line] (4.center)
			 to (2.center)
			 to (5.center)
			 to (6.center)
			 to cycle;
		\draw [style=Line] (5.center) to (3.center);
		\draw [style=Line] (3.center) to (6.center);
		\draw [style=Line] (6.center) to (1.center);
		\draw [style=Line] (1.center) to (4.center);
		\draw [style=Filled line] (8.center)
			 to (9.center)
			 to (10.center)
			 to (7.center)
			 to cycle;
	\end{pgfonlayer}
\end{tikzpicture}

\end{equation}
Note that the skinny zero actually implies the Adler zero (since setting $s_{1n}=0$ further sends $p_1=0$). The goal of this letter is to find the relationship between these hidden zeros and the BCJ relations, as both Tr($\phi^3$) and NLSM satisfy both constraints. We find that all zeros can be derived from BCJ relations for any number of points. Analogous to the soft bootstrap, we use hidden zeros as a constraint, construct higher-derivative corrections to NLSM and compare it to the Adler zero and BCJ relations. Finally, we show that via the KLT relations the hidden zeros carry over to permutationally invariant theories. We show this explicitly in the case of the special Galileon and check both the existence of hidden zeros (now for all permutations) as well as factorizations near these zeros. This raises the question of whether the special Galileon (and other theories produced by the KLT relations) have their origin in a surfacehedron picture \cite{Arkani-Hamed:2023lbd,Arkani-Hamed:2023mvg,Arkani-Hamed:2024vna}.

While this work was in its final stages, the preprint \cite{Cao:2024uni} appeared with some overlapping ideas.

\section{Zeros from BCJ relations}
Let us investigate how the NLSM amplitude zeros are related to the BCJ relations \cite{Bern:2008qj,Bern:2019prr}. At six points the relations can be written in the following form:
\begin{align}
       &A_6[123456]=\frac{1}{s_{12}s_{123}s_{56}}\big[A_6[162543]\  s_{13}s_{25}(s_{56}-s_{24})\nonumber\\&+A_6[162345]\  s_{15}(s_{12}+s_{23})(s_{14}-s_{56})\nonumber\\
    & - A_6[162354]\  s_{14}(s_{12}+s_{23})(s_{25}+s_{35})\label{eq:BCJ6}\\
    &+ A_6[162435]\  s_{13} s_{15}s_{24}
    + A_6[162453]\  s_{13}s_{24}(s_{15}+s_{35}) \nonumber\\
    &- A_6[162534]\  s_{14} s_{25}(s_{12}+s_{23}) \big]
    \,.\nonumber
\end{align}
Since the NLSM has no three-point interactions, its amplitudes have no two-particle poles. Looking at the RHS of \eqref{eq:BCJ6}, it vanishes on both the $C_1$ and $C_2$ zeros defined in \eqref{eq:constrCm}. Thus at six points, the zeros of the NLSM amplitude are a direct consequence of their BCJ-compatibility. 

For Tr($\phi^3$)-theory the relation \eqref{eq:BCJ6} holds when replacing $A_6[\alpha]\to m_6[\alpha|\beta]$ where $m_6$ is a bi-adjoint scalar six-point amplitude and $\beta$ is any ordering such that none of the BCJ basis elements $m_6[162\sigma(345)|\beta]$ vanish. On choosing $\beta\!=\!\{123456\}$ for example on the LHS, none of the amplitudes on the RHS contain poles at the locations of the zeros and so once again the RHS vanishes due to the Mandelstam coefficients. Other orderings vanish trivially through appropriately relabeled BCJ relations. 

The proof that BCJ-compatibility implies the presence of zeros at $n$ points is involved but conceptually straightforward. It requires careful consideration of which terms in the BCJ relation vanish on the support of a generic zero $C_m$ (\ref{eq:constrCm}). Below, we present the details of the proof, starting with a convenient representation of the BCJ relation \cite{Bern:2008qj},
\begin{align}
    A_n[123\cdots n] = (-1)^n  \!\!\!\!\!\sum_{\sigma(3\ldots n-1)}\!\!\!\!  A_n[1n2\sigma]\, \prod_{k=3}^{n-1} \frac{\mathcal{F}_k[2\sigma1]}{s_{kk+1\ldots n}}\label{eq:BCJn}.
\end{align}
The factors $\mathcal{F}_k$ are given by
\begin{equation}
         \mathcal{F}_{k}[\rho] =
    \begin{cases}
        \sum\limits_{l=t_k}^{n-1}\mathcal{S}_{k,\rho_l}+\theta(t_{k-1},t_k) s_{kk+1\ldots n}  &   \text{if } t_{k}>t_{k+1},\!\\
        -\!\!\sum\limits_{l=1}^{t_k}\mathcal{S}_{k,\rho_l} \! -\theta(t_{k},t_{k-1}) s_{kk+1\ldots n}      &   \text{if } t_{k}<t_{k+1},\!
    \end{cases}
\end{equation}    
where $t_k$ is the position of leg $k$ in the ordered list $\rho\!=\!\{2\sigma 1\}$, $\rho_l$ denotes the $l$-th element of $\{2\sigma 1\}$ and we always set $t_2 \equiv 0$ and $t_{n}\equiv t_{n-2}$. The function $\theta(t_i,t_j)=1$ if $t_i >t_j$ and 0 otherwise. The factors $\mathcal{S}_{i,j}$ are defined as
\begin{align}
    \mathcal{S}_{i,j} =
    \begin{cases}
        s_{ij}  &   \text{if } i>j \text{ or } j=1,2,\\
        0       &   \text{else}.
    \end{cases}
\end{align}
Our objective is now to show that for each of the $(n-3)!$ permutations $\{2\sigma 1\}$ where $\sigma \!=\! \sigma(34\ldots n\!-\!1)$ there is at least one index $k\!\in\!\{34\ldots n\!-\!1\}$ such that on the zero constraint $C_m$ (\ref{eq:constrCm}) the factor $\mathcal{F}_k[2\sigma 1]\big\rvert_{C_m} \!\!\!\!= 0$. First we define the set of labels
\begin{align}
    I_m=\{m\!+\!2\ldots n\!-\!1\}.
    \label{Im}
\end{align}
For any term in the BCJ relation corresponding to a particular permutation $\{2\sigma 1\}$ we proceed as follows:
\begin{itemize}
    \item Find the position $t_{m+1}$ of the label $m\!+\!1$ in $\{2\sigma 1\}$.
    \item Identify the label in $I_m$ which appears as the \textit{left-most} in $\{2\sigma 1\}$. Denote it $L$ and its position $t_L$. Note that by definition $t_L<t_i$ for any $i\in I_m\setminus \{L\}\!=\!\{m \!+\!2\ldots\cancel{L}\ldots n\!-\!1\}$.
    \item Identify the label in $I_m$ which appears as the \textit{right-most} in $\{2\sigma 1\}$. Denote it $R$ and its position $t_R$. Note that by definition $t_R>t_i$ for any $i\in I_m\setminus \{R\}\!=\!\{m\!+\!2\ldots\cancel{R}\ldots n\!-\!1\}$.
\end{itemize}
Having carried out this procedure, the permutation $\{2\sigma 1\}$ necessarily falls into one of three categories:
\begin{enumerate}
    \item[(a)] $\{2\sigma 1\} = \{2\ldots m\!+\!1\ldots L \ldots R \ldots 1 \}$, where $t_R>t_{m+1}$ and $t_L>t_{m+1}$,
    \item[(b)] $\{2\sigma 1\} = \{2\ldots  L \ldots R \ldots m\!+\!1 \ldots 1 \}$, where $t_R<t_{m+1}$ and $t_L<t_{m+1}$.
    \item[(c)] $\{2\sigma 1\} = \{2\ldots  L \ldots m\!+\!1 \ldots R  \ldots 1 \}$, where $t_R>t_{m+1}$ and $t_L<t_{m+1}$.
\end{enumerate}
For the case (a) we set $k\!=\!R$. Then by definition of $R$, all elements in $I_m\setminus \{R\}$ appear to the \textit{left} of $R$ in $\{2\sigma 1\}$ and $m+1$ appears to the \textit{left} of $R$. This guarantees that $t_R>t_{R+1}$ and $t_R>t_{R-1}$ for any $R\in\! I_m$, leading to
\begin{align}
    (\text{a})\!:\,\,\, \mathcal{F}_{R}[2\sigma 1] = \sum_{i=t_R}^{n-1} \mathcal{S}_{R,\rho_i} = \!\!\sum_{R > \rho_i} s_{R\rho_i}.
\end{align}
Since we are summing over all labels to the \textit{right} of $R$, the range for $\rho_i$ can only include $\rho_i \subset \{134\ldots m\}$. But for any choice of $R\in\! I_m$ and $\rho_i \in \{134\ldots m\}$, the invariant $s_{R\rho_i}$ will be part of the constraint $C_m$ (\ref{eq:constrCm}). Hence
\begin{align}
    (\text{a})\!:\,\,\, \mathcal{F}_{R}[2\sigma 1]\big\rvert_{C_m} \!\!\!\! = 0.
\end{align}
Setting $k=L$ for case (b) and either $k=L$ or $k=R$ for case (c), one can show by analogous arguments that 
\begin{align}
    (\text{b})\!:\,\,\, \mathcal{F}_{L}[2\sigma 1]\big\rvert_{C_m} \!\!\!\! = 0,
    \hspace{0.5cm}
    (\text{c})\!:\,\,\, \mathcal{F}_{R/L}[2\sigma 1]\big\rvert_{C_m} \!\!\!\! = 0.
\end{align}
Thus for any permutation $\{2\sigma 1\}$ appearing in the BCJ relation (\ref{eq:BCJn}), there exists at least one $k\!\in\!I_m$ such that $\mathcal{F}_k[2\sigma 1]\big\rvert_{C_m} \!\!\!\!= 0$  holds. This in turn proves that 
\begin{align}
 A_n[12\ldots n]\big\rvert_{C_m} \!\!\!\!= 0,
\end{align}
for any $C_m$ of the form (\ref{eq:constrCm}).
\section{Zeros of higher-derivative theories}

Following the logic of the soft and BCJ bootstraps \cite{Low:2019ynd, Brown:2023srz, Elvang:2018dco, Kampf:2021bet,Kampf:2021jvf} we can now ask about the existence of zeros for higher-derivative corrections to NLSM theory, and compare the constraints they impose to the Adler zero conditions and BCJ relations. Let us look at the first generic case of six-point amplitudes. At the two-derivative or ${\cal O}(p^2)$ level, we find that the NLSM amplitude is the unique solution to all three questions: it is completely fixed by the Adler zero, it is the only BCJ-satisfying amplitude, and it is the unique solution to the hidden zero problem. We start with a general ansatz for the six-point amplitude in ${\cal O}(p^{2k})$ power-counting, 
\begin{equation}
    A_6 = \sum_i \sum_{\rm cycl} \frac{a_i P_i}{s_{123}} + \sum_j b_j Q_j \,,\label{ans6pt}
\end{equation}
where $P_i\sim s^{k{+}1}$, $Q_j\sim s^k$ are polynomials in $s_{ij}$. Imposing the zero conditions $C_1$ and $C_2$ allows us to constrain the coefficients $a_i$, $b_j$.
For $k=1$ all coefficients get fixed and we find a unique solution -- the NLSM amplitude,
\begin{equation}
    A_6^{\rm NLSM} = -\frac{1}{2} \frac{(s_{12}+s_{23})(s_{45}+s_{56})}{s_{123}}+ s_{12} + \text{cyc}.
\end{equation}
For $k{=}2$, i.e. at ${\cal O}(p^4)$ order, there are no BCJ-satisfying six-point (and higher-point) amplitudes, but starting with (\ref{ans6pt}) and applying the zero conditions (\ref{6pC}), we find one solution.
It can be identified as the amplitude of a particular operator in the chiral Lagrangian (with coefficient $L_0$ in the conventions of \cite{Bijnens:2019eze}). Explicitly, it reads
\begin{align}
    A_6^{(4)}  &= s_{12}(s_{12}+s_{34}+2s_{45})-s_{123}(s_{345}-2s_{61})\nonumber\\
    &\hspace{2cm}-\frac{(s_{12}+s_{23})(s_{45}+s_{56})^2}{s_{123}}+\text{cyc}. 
    \label{A6op4}
\end{align}
On poles, this consistently factorizes into an NLSM amplitude and a higher-derivative four-point amplitude,
\begin{align}
    A_{4}^{(4)} = (s_{12}+s_{23})^2 = s_{13}^2,
    \label{A4op4}
\end{align}
which itself satisfies the four-point zero conditions.

Interestingly, the BCJ-incompatible amplitude (\ref{A6op4}) factorizes close to the zeros, similar to the leading order NLSM. Let us first consider the skinny zero  $C_1$ in (\ref{6pC}). Turning on $s^*=s_{15}\ne 0$ in (\ref{Afact}) we find,
\begin{align}
  \hspace{-0.5cm} A_{6}^{(4)}\big\rvert_{C_1} \xrightarrow{s_{15}\neq 0}  A_{4}^{(2)}{\times} m_{3,B}^{(0)}{\times} A_{5,T}^{(2)} + A_{4}^{(4)}{\times} m_{3,B}^{(0)} {\times} A_{5,T}^{(0)}\,,
  \label{C1c15}
\end{align}
where $m_{3,B}^{(0)}=1$ is the three-point Tr($\phi^3$) amplitude and we identify (as functions of $s_{ij}$) the pre-factors
\begin{align}
\begin{split}
   A_{4}^{(2)} = -s_{15}, \hspace{0.5cm} A_{4}^{(4)} = s_{15}^2,
   \label{4ptFct}
\end{split}
\end{align}
as the NLSM amplitude and its hidden zero-compliant higher-derivative correction (\ref{A4op4}). The five-point ``top'' amplitudes in (\ref{C1c15}) involve pions and Tr($\phi^3$) scalars,
\begin{align}
    A_{5,T}^{(0)}(\phi\pi\pi\phi\phi)  = 1 - \frac{s_{23}+s_{34}}{s_{234}}-\frac{s_{34}+s_{45}}{s_{345}},
\end{align}
as well as the higher-derivative correction,
\begin{align}
    A_{5,T}^{(2)}(\phi\pi\pi\phi\phi) {=} s_{23} {+} 2s_{34} {+} s_{45} {-} \frac{(s_{23}{+}s_{34})^2}{s_{234}}{-}\frac{(s_{34}{+}s_{45})^2}{s_{345}}.
\end{align}
In particular the higher-derivative amplitude retains the Adler zero for pions and correctly factorizes on poles into (\ref{A4op4}) and the three-scalar amplitude $m_{3,B}$.
The factorization (\ref{C1c15}) presents a natural generalization of the known factorization theorem for NLSM amplitudes \cite{Arkani-Hamed:2023swr}. In an EFT expansion, we get a sum over products of NLSM and mixed amplitudes (with higher-derivative corrections) such that their combined mass dimension matches the given power counting.
Setting $s^*$ to a different $s_{ij}$ in (\ref{C1c15}) leads to the appearance of different mixed amplitudes and their higher-derivative corrections. The distribution of pions and scalars follows the pattern described for the leading order NLSM in \cite{Arkani-Hamed:2023swr}.

The even-even factorization for higher-derivative amplitudes close to the zero $C_2$ in (\ref{6pC}) also generalizes the behavior of NLSM amplitudes. For $s^*=s_{25}$,
\begin{align}
  A_{6}^{(4)}\big\rvert_{C_2} \xrightarrow{s_{25}\neq 0 }  m_4 \times \big\lbrace A_{4,B}^{(4)}\! \times A_{4,T}^{(2)} {+} A_{4,B}^{(2)}\times A_{4,T}^{(4)}  \big\rbrace,
\end{align}
where the prefactor $m_4$ takes the form of a four-point Tr($\phi^3$) amplitude
\begin{align}
    m_4 = \frac{1}{s_{123}} +\frac{1}{s_{345}},
    \label{m4BAS}
\end{align}
and the ``bottom'' and ``top'' amplitudes only involve (higher-derivative) pions.

Returning to the bootstrap analysis, we note that while the amplitude (\ref{A6op4}) satisfies the Adler zero condition, we want to emphasize that the constraints from the hidden zeros are in fact \emph{more} stringent. At $\mathcal{O}(p^4)$ there are two independent operators in the chiral Lagrangian. Both generate amplitudes with the Adler zero, but only (\ref{A6op4}) also has the hidden zeros. We can continue to higher ${\cal O}(p^{2k})$ orders and find a general pattern,
\begin{equation}
    (\mbox{BCJ satisfying}) \subset (\mbox{Hidden zeros}) \subset (\mbox{Adler zero})
\end{equation}
which can be supported by the explicit counting of solutions to higher $k$ found in table \ref{tab6pt}.
\vspace{-0.3cm}
\begin{table}[H]
\centering
\caption{Number of six-point amplitudes from various constraints in NLSM}
\label{tab6pt}
\vspace{0.2cm}
\begin{tabular}{ |c||c|c|c|c|c|c|c|c|} 
 \hline
 ${\cal O}(p^\#)$ &  2 &  4 &  6 &  8 & 10 & 12 & 14      \\ \hline
Alder zero & 1 & 2 & 10 & 29 & 78 & 203 & 461    \\ \hline
Hidden zeros & 1 & 1 & 5 & 13 & 41 & 112 & 282    \\ \hline
BCJ satisfying & 1 & 0 & 1 & 1 & 2 & 4 & 7      \\ 
 \hline
\end{tabular}
\end{table}
As a fun side check, one can repeat the exercise imposing only skinny zeros $C_1$ (ignoring the rectangle zero $C_2$). Here we get more solutions. In particular, at ${\cal O}(p^8)$ order, we get 14 solutions which reduce to 13 when we impose the rectangle zero. This suggests a pattern that the presence of skinny zeros are stronger conditions than that of Adler zeros, and rectangle zeros provide further constraints when fixing coefficients in the ansatz (\ref{ans6pt}).

\section{KLT relations and special Galileon}

The geometric origin of hidden zeros \cite{Arkani-Hamed:2023swr} suggests that their presence requires an ordered set of external states. On the other hand, since the BCJ relations imply their existence (assuming the absence of two-particle poles), a natural question is whether or not these zeros persist in permutation-invariant double copy theories. In particular: do the hidden zeros in the NLSM double copy to zeros of tree amplitudes in special Galileon theory? Note that Tr($\phi^3$) is not a BCJ-compatible theory, nor is it the product of a double copy and so the only scalar theory of interest here will be the special Galileon theory \cite{Bern:2019prr}.
	
Let us start with the six-point amplitude,
	\begin{align}
		\label{eq:sGal6}
		M_6=\frac{s_{12}s_{23}s_{13}s_{45}s_{46}s_{56}}{s_{123}}+\text{perms}-\frac12G(12345)\,,
	\end{align}
where $G$ denotes the Gram determinant. On the zero $C_1$, all terms but the following vanish
	\begin{align}
		\frac{s_{12}s_{26}s_{16}s_{34}s_{45}s_{35}}{s_{126}}+ s_{12}^2 s_{34}s_{35}s_{45}\,.
	\end{align}
On the support of $C_1$, $s_{12}=-s_{16}$ and we see that the above expression must vanish -- $C_1$ is indeed a zero of the special Galileon amplitude! Indeed the amplitude \eqref{eq:sGal6} vanishes on $C_2$ as well and the permutation symmetry of the amplitude means that it vanishes at many more locations than its single copy. 
	
Moving on to eight points, we find that the Galileon amplitude vanishes on all three expected zeros $C_1$, $C_2$ and $C_3$. In the next section, we will see that these zeros are not a low-point accident, but rather their presence is guaranteed by the KLT double copy formula. 
	
In light of the known factorization theorem (\ref{Afact}) for ordered theories, let us now study the behavior of special Galileon amplitudes close to the hidden zeroes.
The structure of the known theorem is determined by the color-ordering of external states, making it seem unlikely that permutation-invariant double copy products inherit such properties. Contrary to this expectation, we find that special Galileon amplitudes indeed do factorize near their zeros. Take for example the six-point amplitude \eqref{eq:sGal6} near the zero $C_2$,
	\begin{align}
		M_6\big\rvert_{C_2} \xrightarrow{s_{25}\neq 0} m_4 \times M_{4,B} \times M_{4,T}\,,
	\end{align}
where $m_4$ is as in (\ref{m4BAS}) and the ``top''- and ``bottom''-amplitude can be functionally identified as the four-point special Galileon,
\begin{align}
    M_{4,B} = s_{61}s_{12}(s_{61}+s_{12}), \hspace{0.3cm} M_{4,T} = s_{34}s_{45}(s_{34}+s_{45}).
\end{align}
Thus, despite the permutation invariance of Galileon theory, it still satisfies the same factorization theorem as NLSM and Tr($\phi^3$). We checked that this factorization near zeros persists at eight points. 

We end this section with a short discussion on factorization into odd-point subamplitudes. Already in the case of NLSM, this involves the introduction of mixed pion-scalar amplitudes in a theory of NLSM coupled to a Tr($\phi^3$) scalar $\phi$. Due to its analogous role in the soft limit and double copy, a natural guess for odd-point factorization in Galileon theory is subamplitudes in a theory of a Galileon coupled to pions and scalars, first studied in \cite{Cachazo:2016njl}. For example near $C_1$ with $s_{15}\ne 0$,
\begin{align}
    M_6\big\rvert_{C_1} \xrightarrow{s_{15}\neq 0} M_4 \times \tilde{M}_{5,T}
\end{align}
where $\tilde{M}_{5,T}\!=\!\tilde{M}_{5}(\phi\, \Gamma\,\Gamma\, \phi\,\phi)$ involves special Galileons and scalars. Limited checks at six-point verify this guess, but a deeper discussion is left for future work.
 
\section{KLT preserves zeros}
Encouraged by the explicit checks of the previous section at six and eight points, we investigate how the special Galileon amplitudes inherit the zeros through the double copy procedure.
To see how this works, we can write the KLT kernel in a basis where each element vanishes at these zeros, which implies that the KLT formula,
\begin{equation}
\label{eq:KLTformula}
    M_n = \sum_{\sigma \gamma} S[\sigma | \gamma] A_n(1 \gamma m{+}1 n) A_n(1 m{+}1 n \sigma)\,,
\end{equation}
preserves the zeros $C_m$ if the double copied amplitude has no two-particle poles.
\par
For a given zero $C_m$ as in (\ref{eq:constrCm}), let us consider the two sets $I_m$, as defined in (\ref{Im}), and $J_m \!=\!\{2,3,...,m \}$ such that $s_{zf}=0$ for any $z \!\in\! I_m$ and $f \!\in\! J_m$.  The form of the KLT kernel \cite{Kawai:1985xq, Bjerrum-Bohr:2010pnr} we will be using is given by 
\begin{align}
\label{eq:kernel}
    &S[a_1,\cdots a_{n-3}|b_1\cdots b_{n-3}]\\
&\hspace{0.5cm}=\left(\frac{1}2\right)^{-(n-3)}\prod_{t=1}^{n-3} \left(p_1\cdot p_{a_t}+\sum_{q>t}^{n-3}\theta(a_{t},a_{q})p_{a_t}\cdot p_{a_q}\right)\!, \nonumber
\end{align}
and is evaluated in the KLT formula on different permutations i.e. $S[\sigma(a_1,\cdots a_{n-3})|\gamma(b_1,\cdots, b_{n-3})]$. It is convenient to choose a basis for the kernel where $\gamma$ and $\sigma$ contain only elements that live in $J_m$ or $I_m$. This corresponds to fixing the positions of $\{1, m{+}1, n\}$ as in \eqref{eq:KLTformula}, instead of the usual choice of $\{1, n{-}1, n\}$ \cite{Kawai:1985xq,Bjerrum-Bohr:2010pnr}. Note that if $s_{ij}=0$ is a zero of $M(12...m{+}1...n{-}1 \ n)$, it is also a zero of $M(12...n{-}1...m{+}1 \ n)$ due to the permutational invariance of $M$, allowing a $m{+}1\leftrightarrow n{-}1$ swap in \eqref{eq:KLTformula}.

The KLT kernel is proportional to $s_{1 z_0}\!=\!0$ for any $z_0 \!\in\! I_m$ if the permutation $\sigma$ ends with $z_0$. Similarly, if $\gamma$ starts with any $ z_1 \in I_m$, the kernel is zero. This is because as soon as we consider $i_t = z_1$ in the kernel \eqref{eq:kernel}, anything that follows $z_1$ in the $\sigma$ ordering necessarily follows it in the $\gamma$ ordering. This preserves the order in both sets and sets the KLT kernel proportional to $s_{1 z_1}=0$. Thus the only terms of interest in the KLT formula are of the following form:
\begin{equation}
    M = \sum_{\sigma \gamma} S[\sigma f_0 | f_1 \gamma] A_n(1 f_1 \gamma m+1 n) A_n(1 f_0 \Bar{\sigma} n m+1)
\end{equation}
where $f_0, f_1 \in J_m$ and $\Bar{\sigma}$ denotes the reversed $\sigma$ ordering.  The non-zero contributions to $M$ are then orderings where $\gamma$ starts with an element of $J_m$ and $\sigma$ ends with one. Denoting the rightmost label in $\sigma$ that lives in $I_m$ as $z_0$, the kernel is proportional to $s_{1z_0}+\sum_{f > z_0} s_{z_0 f}$. Here the sum is over all labels $f$ in $\sigma$ to the right of $z_0$, such that $f$ is to the left of $z_0$ in $\gamma$. However, if $z_0 \in I_m$ and $f \in J_m$, $s_{z_0 f} = 0$, meaning the kernel vanishes. As a result, all the terms contributing to $M$ vanish, and the KLT formula preserves the zeros $C_m$ as long as the amplitude does not have any two-particle poles.

Unlike the presence of zeros in double copy theories, factorization near them is less apparent from the KLT formula. We turned instead to a different formulation of Galileon amplitudes, the CHY formula \cite{Cachazo:2014xea}. Though less straightforward, it is possible to see that the factorization of Galileon tree amplitudes into lower-point ones near ``even'' zeros $C_{2m}$ is a consequence of the CHY formula, as we will discuss further in upcoming work \cite{inprogress:Bartsch}. 

\section{Zeros in special Galileon EFT}

In this section, we investigate the hidden zeros from the point of view of higher-derivative corrections to the leading order NLSM and special Galileon amplitudes. It was shown in \cite{Elvang:2018dco, Brown:2023srz, CarrilloGonzalez:2019fzc} that certain special higher-derivative corrections to NLSM satisfy BCJ relations and results were provided up to ${\cal O}(p^{18})$ order for six-point amplitudes. In general, the BCJ conditions on higher-point amplitudes provided constraints on lower-point amplitudes, and it is an open question how many BCJ satisfying theories there are to all points. 

The BCJ-satisfying ordered amplitudes can be plugged into the KLT formula, producing amplitudes in a higher-derivative single scalar theory. In the last section, we saw that the KLT formula preserves all zeros of the ordered amplitude -- in fact, the new amplitude has hidden zeros at all permutations of the kinematic conditions (\ref{eq:constrCm}). In \cite{Brown:2023srz} we inspected these amplitudes from the point of view of soft limits. The BCJ amplitudes have ${\cal O}(t)$ Adler zeros \cite{Adler:1964um} and the KLT relations enhance this behavior to ${\cal O}(t^3)$, as evident from the NLSM - special Galileon examples. Though this works more generally, we find that not all ${\cal O}(t^3)$ amplitudes are generated from KLT (the first counter example appears at 16-derivative order), only a special subset is. Here we explore this question in the context of hidden zeros. In the schematic KLT formula,
\begin{equation}
    (\mbox{KLT amp}) = (\mbox{BCJ amp})\otimes(\mbox{BCJ amp})
\end{equation}
we can count the number of possible terms on the left hand side and compare them to the scalar bootstrap construction of all amplitudes with hidden zeros at a given derivative order. Because the hidden zero conditions are trivial for four-point amplitudes, we begin the analysis at six points. In particular, we find:

\begin{table}[ht]
\centering
\caption{Number of six-point amplitudes from various constraints in special Galileon theory}
\label{tab6pt:sGal}
\begin{tabular}{ |c||c|c|c|c|c|c|c|c|} 
 \hline
 ${\cal O}(p^\#)$ &  10 &  12 &  14 &  16 & 18 & 20 & 22      \\ \hline
${\cal O}(t^3)$ soft behavior & 1 & 0 & 1 & 3 & 10 & 23 & 49    \\ \hline
Hidden zeros & 1 & 0 & 1 & 1 & 4 & 6 & 14    \\ \hline
Generated from KLT & 1 & 0 & 1 & 1 & 3 & 5 & 10      \\ 
 \hline
\end{tabular}
\end{table}
At ${\cal O}(p^{10})$ order, we identify one solution for each of the constraints and that is the special Galileon. There is no solution at the next ${\cal O}(p^{12})$ order to any of the constraints, and one for each at the ${\cal O}(p^{14})$ order (all corresponding to the same theory). The first difference appears at ${\cal O}(p^{16})$, where the ${\cal O}(t^3)$ soft limit constraints produce three solutions while only one of them is generated from the KLT relations and also enjoys the hidden zeros. Note that up to this derivative order one can identify all solutions to the ${\cal O}(t^3)$ constraint as higher-derivative corrections to the special Galileon theory. At the next level, ${\cal O}(p^{18})$ we find that out of the four solutions to the hidden zeros constraint, only three of them are generated by the KLT relations. This leaves us with the following hierarchy of constraints,
\begin{equation*}
    (\mbox{From KLT}) \subset (\mbox{Hidden zeros}) \subset ({\cal O}(t^3)\,\mbox{behavior})
\end{equation*}
Note that here we only imposed the presence of hidden zeros as a constraint, not factorization near these zeros. To do that one would need to provide a precise prescription for the higher-derivative theories. It is possible that on doing so, this restricted set of ``zeros+factorization'' theories is the same as the set of all ``KLT theories''. 

Here it is important to note that the factorization theorem for EFT amplitudes is modified when compared to the leading order NLSM or Galileon theories. They now involve a higher-derivative Tr($\phi^3$) amplitude, leading to
\begin{align}
    M^\text{h.d.}_n\big\rvert_{C_m} \xrightarrow[]{s^\ast \neq 0} A^{\text{Tr}\phi^3\text{+h.d.}}_4 \times M^\text{h.d.}_B \times M^\text{h.d.}_T\,.
\end{align}
The Tr($\phi^3$) amplitude with higher-derivative corrections obtained from the near-zero factorization of both BCJ-compatible NLSM and special Galileon theory are equal,
\begin{align}
\label{eq:BAShd}
    \hspace{-0.25cm} A^{\text{Tr}\phi^3\text{+h.d.}}_4\!=\!\Lambda\! \left(\frac{1}{s}{+}\frac{1}{t}\right) \!{-} \frac{t}{\Lambda} {+}\frac{t^2}{\Lambda^2} {+}\frac{t}{\Lambda^3} (7 t^2 {-} us)+\cdots
\end{align}
where $\Lambda$ is the EFT scale. Tr($\phi^3$) amplitudes coincide with diagonal bi-adjoint scalar amplitudes, whose higher-derivative corrections have been studied in the context of modified KLT kernels \cite{Chi:2021mio, Chen:2023dcx, Chen:2022shl,Carrasco:2019yyn, Carrasco:2021ptp} and \eqref{eq:BAShd} is one with a specific choice of Wilson coefficients. What type of KLT kernel this leads to and whether we can use a modified kernel to make the factorization theorem for leading and subleading orders coincide are interesting questions left for future work.

\section{Outlook}
The relationship between hidden zeros and the double copy discussed above opens many avenues of investigation. Some, such as the possible relation between higher-derivative KLT kernels and factorization near zeros of EFT amplitudes, as well as a detailed study of Galileon-pion-scalar mixed amplitudes were discussed in the text above. The most interesting one relates to the presence of zeros and factorization in gravitational amplitudes obtained as a double copy from YM amplitudes. The YM zeros as described in \cite{Arkani-Hamed:2023swr} include conditions involving the polarization vectors as well as momenta. Additionally the amplitudes contain 2-particle poles, voiding the KLT argument. Nonetheless, we find that special properties of YM ensure that gravity amplitudes do indeed vanish on these zeros via the CHY formula, which we discuss in an upcoming work \cite{inprogress:Bartsch}.

The original context of the hidden zeros in NLSM and Tr($\phi^3$) was that of the associahedron \cite{nima2, Arkani-Hamed:2019vag, Arkani-Hamed:2017mur}, but an independent geometric description of NLSM, similar to the amplituhedron \cite{Arkani-Hamed:2013jha,Arkani-Hamed:2013kca,Arkani-Hamed:2017vfh,Damgaard:2019ztj,He:2023rou} for planar ${\cal N}=4$ SYM or the Tr($\phi^3$) associahedron itself is yet-to-be found. Furthermore, nothing is known about the geometric description of permutation-invariant amplitudes and any insight into the geometric origin of Galileon zeros is absent. 

A final interesting question about these hidden zeros is that of uniqueness of tree-level amplitudes \cite{Arkani-Hamed:2016rak,Rodina:2016jyz,Rodina:2016mbk} and also loop integrands \cite{Arkani-Hamed:2010zjl,Bartsch:2022pyi,Bartsch:2024ofb}. For the NLSM, one can write down a recursion relation similar to soft subtracted recursion \cite{Kampf:2012fn,Cheung:2015ota, Luo:2015tat,Cheung:2016drk,Kampf:2019mcd}. To what extent double copy theories are fixed by the presence of their zeros, especially in an EFT expansion is an open problem.

\medskip

{\it Acknowledgements}: We thank Nima Arkani-Hamed, Carolina Figueiredo, Song He, Alfredo Guevara, Callum Jones and Jiri Novotny for very useful discussions. This work is supported by GA\-\v{C}R 24-11722S, GAUK-327422, MEYS LUAUS23126, DOE grant No. SC0009999 and the funds of the University of California.
\bibliography{mainbib}

\begin{thebibliography}{56}%
\makeatletter
\providecommand \@ifxundefined [1]{%
 \@ifx{#1\undefined}
}%
\providecommand \@ifnum [1]{%
 \ifnum #1\expandafter \@firstoftwo
 \else \expandafter \@secondoftwo
 \fi
}%
\providecommand \@ifx [1]{%
 \ifx #1\expandafter \@firstoftwo
 \else \expandafter \@secondoftwo
 \fi
}%
\providecommand \natexlab [1]{#1}%
\providecommand \enquote  [1]{``#1''}%
\providecommand \bibnamefont  [1]{#1}%
\providecommand \bibfnamefont [1]{#1}%
\providecommand \citenamefont [1]{#1}%
\providecommand \href@noop [0]{\@secondoftwo}%
\providecommand \href [0]{\begingroup \@sanitize@url \@href}%
\providecommand \@href[1]{\@@startlink{#1}\@@href}%
\providecommand \@@href[1]{\endgroup#1\@@endlink}%
\providecommand \@sanitize@url [0]{\catcode `\\12\catcode `\$12\catcode `\&12\catcode `\#12\catcode `\^12\catcode `\_12\catcode `\%12\relax}%
\providecommand \@@startlink[1]{}%
\providecommand \@@endlink[0]{}%
\providecommand \url  [0]{\begingroup\@sanitize@url \@url }%
\providecommand \@url [1]{\endgroup\@href {#1}{\urlprefix }}%
\providecommand \urlprefix  [0]{URL }%
\providecommand \Eprint [0]{\href }%
\providecommand \doibase [0]{http://dx.doi.org/}%
\providecommand \selectlanguage [0]{\@gobble}%
\providecommand \bibinfo  [0]{\@secondoftwo}%
\providecommand \bibfield  [0]{\@secondoftwo}%
\providecommand \translation [1]{[#1]}%
\providecommand \BibitemOpen [0]{}%
\providecommand \bibitemStop [0]{}%
\providecommand \bibitemNoStop [0]{.\EOS\space}%
\providecommand \EOS [0]{\spacefactor3000\relax}%
\providecommand \BibitemShut  [1]{\csname bibitem#1\endcsname}%
\let\auto@bib@innerbib\@empty
\bibitem [{\citenamefont {Britto}\ \emph {et~al.}(2005)\citenamefont {Britto}, \citenamefont {Cachazo}, \citenamefont {Feng},\ and\ \citenamefont {Witten}}]{Britto:2005fq}%
  \BibitemOpen
  \bibfield  {author} {\bibinfo {author} {\bibfnamefont {R.}~\bibnamefont {Britto}}, \bibinfo {author} {\bibfnamefont {F.}~\bibnamefont {Cachazo}}, \bibinfo {author} {\bibfnamefont {B.}~\bibnamefont {Feng}}, \ and\ \bibinfo {author} {\bibfnamefont {E.}~\bibnamefont {Witten}},\ }\href {\doibase 10.1103/PhysRevLett.94.181602} {\bibfield  {journal} {\bibinfo  {journal} {Phys. Rev. Lett.}\ }\textbf {\bibinfo {volume} {94}},\ \bibinfo {pages} {181602} (\bibinfo {year} {2005})},\ \Eprint {http://arxiv.org/abs/hep-th/0501052} {arXiv:hep-th/0501052} \BibitemShut {NoStop}%
\bibitem [{\citenamefont {Cohen}\ \emph {et~al.}(2011)\citenamefont {Cohen}, \citenamefont {Elvang},\ and\ \citenamefont {Kiermaier}}]{Cohen:2010mi}%
  \BibitemOpen
  \bibfield  {author} {\bibinfo {author} {\bibfnamefont {T.}~\bibnamefont {Cohen}}, \bibinfo {author} {\bibfnamefont {H.}~\bibnamefont {Elvang}}, \ and\ \bibinfo {author} {\bibfnamefont {M.}~\bibnamefont {Kiermaier}},\ }\href {\doibase 10.1007/JHEP04(2011)053} {\bibfield  {journal} {\bibinfo  {journal} {JHEP}\ }\textbf {\bibinfo {volume} {04}},\ \bibinfo {pages} {053} (\bibinfo {year} {2011})},\ \Eprint {http://arxiv.org/abs/1010.0257} {arXiv:1010.0257 [hep-th]} \BibitemShut {NoStop}%
\bibitem [{\citenamefont {Cheung}\ \emph {et~al.}(2015{\natexlab{a}})\citenamefont {Cheung}, \citenamefont {Shen},\ and\ \citenamefont {Trnka}}]{Cheung:2015cba}%
  \BibitemOpen
  \bibfield  {author} {\bibinfo {author} {\bibfnamefont {C.}~\bibnamefont {Cheung}}, \bibinfo {author} {\bibfnamefont {C.-H.}\ \bibnamefont {Shen}}, \ and\ \bibinfo {author} {\bibfnamefont {J.}~\bibnamefont {Trnka}},\ }\href {\doibase 10.1007/JHEP06(2015)118} {\bibfield  {journal} {\bibinfo  {journal} {JHEP}\ }\textbf {\bibinfo {volume} {06}},\ \bibinfo {pages} {118} (\bibinfo {year} {2015}{\natexlab{a}})},\ \Eprint {http://arxiv.org/abs/1502.05057} {arXiv:1502.05057 [hep-th]} \BibitemShut {NoStop}%
\bibitem [{\citenamefont {Adler}(1965)}]{Adler:1964um}%
  \BibitemOpen
  \bibfield  {author} {\bibinfo {author} {\bibfnamefont {S.~L.}\ \bibnamefont {Adler}},\ }\href {\doibase 10.1103/PhysRev.137.B1022} {\bibfield  {journal} {\bibinfo  {journal} {Phys. Rev.}\ }\textbf {\bibinfo {volume} {137}},\ \bibinfo {pages} {B1022} (\bibinfo {year} {1965})}\BibitemShut {NoStop}%
\bibitem [{\citenamefont {Kampf}\ \emph {et~al.}(2013{\natexlab{a}})\citenamefont {Kampf}, \citenamefont {Novotny},\ and\ \citenamefont {Trnka}}]{Kampf:2013vha}%
  \BibitemOpen
  \bibfield  {author} {\bibinfo {author} {\bibfnamefont {K.}~\bibnamefont {Kampf}}, \bibinfo {author} {\bibfnamefont {J.}~\bibnamefont {Novotny}}, \ and\ \bibinfo {author} {\bibfnamefont {J.}~\bibnamefont {Trnka}},\ }\href {\doibase 10.1007/JHEP05(2013)032} {\bibfield  {journal} {\bibinfo  {journal} {JHEP}\ }\textbf {\bibinfo {volume} {05}},\ \bibinfo {pages} {032} (\bibinfo {year} {2013}{\natexlab{a}})},\ \Eprint {http://arxiv.org/abs/1304.3048} {arXiv:1304.3048} \BibitemShut {NoStop}%
\bibitem [{\citenamefont {Cheung}\ \emph {et~al.}(2015{\natexlab{b}})\citenamefont {Cheung}, \citenamefont {Kampf}, \citenamefont {Novotny},\ and\ \citenamefont {Trnka}}]{Cheung:2014dqa}%
  \BibitemOpen
  \bibfield  {author} {\bibinfo {author} {\bibfnamefont {C.}~\bibnamefont {Cheung}}, \bibinfo {author} {\bibfnamefont {K.}~\bibnamefont {Kampf}}, \bibinfo {author} {\bibfnamefont {J.}~\bibnamefont {Novotny}}, \ and\ \bibinfo {author} {\bibfnamefont {J.}~\bibnamefont {Trnka}},\ }\href {\doibase 10.1103/PhysRevLett.114.221602} {\bibfield  {journal} {\bibinfo  {journal} {Phys. Rev. Lett.}\ }\textbf {\bibinfo {volume} {114}},\ \bibinfo {pages} {221602} (\bibinfo {year} {2015}{\natexlab{b}})},\ \Eprint {http://arxiv.org/abs/1412.4095} {arXiv:1412.4095} \BibitemShut {NoStop}%
\bibitem [{\citenamefont {Cheung}\ \emph {et~al.}(2016)\citenamefont {Cheung}, \citenamefont {Kampf}, \citenamefont {Novotny}, \citenamefont {Shen},\ and\ \citenamefont {Trnka}}]{Cheung:2015ota}%
  \BibitemOpen
  \bibfield  {author} {\bibinfo {author} {\bibfnamefont {C.}~\bibnamefont {Cheung}}, \bibinfo {author} {\bibfnamefont {K.}~\bibnamefont {Kampf}}, \bibinfo {author} {\bibfnamefont {J.}~\bibnamefont {Novotny}}, \bibinfo {author} {\bibfnamefont {C.-H.}\ \bibnamefont {Shen}}, \ and\ \bibinfo {author} {\bibfnamefont {J.}~\bibnamefont {Trnka}},\ }\href {\doibase 10.1103/PhysRevLett.116.041601} {\bibfield  {journal} {\bibinfo  {journal} {Phys. Rev. Lett.}\ }\textbf {\bibinfo {volume} {116}},\ \bibinfo {pages} {041601} (\bibinfo {year} {2016})},\ \Eprint {http://arxiv.org/abs/1509.03309} {arXiv:1509.03309} \BibitemShut {NoStop}%
\bibitem [{\citenamefont {Cheung}\ \emph {et~al.}(2017)\citenamefont {Cheung}, \citenamefont {Kampf}, \citenamefont {Novotny}, \citenamefont {Shen},\ and\ \citenamefont {Trnka}}]{Cheung:2016drk}%
  \BibitemOpen
  \bibfield  {author} {\bibinfo {author} {\bibfnamefont {C.}~\bibnamefont {Cheung}}, \bibinfo {author} {\bibfnamefont {K.}~\bibnamefont {Kampf}}, \bibinfo {author} {\bibfnamefont {J.}~\bibnamefont {Novotny}}, \bibinfo {author} {\bibfnamefont {C.-H.}\ \bibnamefont {Shen}}, \ and\ \bibinfo {author} {\bibfnamefont {J.}~\bibnamefont {Trnka}},\ }\href {\doibase 10.1007/JHEP02(2017)020} {\bibfield  {journal} {\bibinfo  {journal} {JHEP}\ }\textbf {\bibinfo {volume} {02}},\ \bibinfo {pages} {020} (\bibinfo {year} {2017})},\ \Eprint {http://arxiv.org/abs/1611.03137} {arXiv:1611.03137} \BibitemShut {NoStop}%
\bibitem [{\citenamefont {Elvang}\ \emph {et~al.}(2019)\citenamefont {Elvang}, \citenamefont {Hadjiantonis}, \citenamefont {Jones},\ and\ \citenamefont {Paranjape}}]{Elvang:2018dco}%
  \BibitemOpen
  \bibfield  {author} {\bibinfo {author} {\bibfnamefont {H.}~\bibnamefont {Elvang}}, \bibinfo {author} {\bibfnamefont {M.}~\bibnamefont {Hadjiantonis}}, \bibinfo {author} {\bibfnamefont {C.~R.~T.}\ \bibnamefont {Jones}}, \ and\ \bibinfo {author} {\bibfnamefont {S.}~\bibnamefont {Paranjape}},\ }\href {\doibase 10.1007/JHEP01(2019)195} {\bibfield  {journal} {\bibinfo  {journal} {JHEP}\ }\textbf {\bibinfo {volume} {01}},\ \bibinfo {pages} {195} (\bibinfo {year} {2019})},\ \Eprint {http://arxiv.org/abs/1806.06079} {arXiv:1806.06079} \BibitemShut {NoStop}%
\bibitem [{\citenamefont {Cachazo}\ \emph {et~al.}(2015)\citenamefont {Cachazo}, \citenamefont {He},\ and\ \citenamefont {Yuan}}]{Cachazo:2014xea}%
  \BibitemOpen
  \bibfield  {author} {\bibinfo {author} {\bibfnamefont {F.}~\bibnamefont {Cachazo}}, \bibinfo {author} {\bibfnamefont {S.}~\bibnamefont {He}}, \ and\ \bibinfo {author} {\bibfnamefont {E.~Y.}\ \bibnamefont {Yuan}},\ }\href {\doibase 10.1007/JHEP07(2015)149} {\bibfield  {journal} {\bibinfo  {journal} {JHEP}\ }\textbf {\bibinfo {volume} {07}},\ \bibinfo {pages} {149} (\bibinfo {year} {2015})},\ \Eprint {http://arxiv.org/abs/1412.3479} {arXiv:1412.3479} \BibitemShut {NoStop}%
\bibitem [{\citenamefont {Casali}\ \emph {et~al.}(2015)\citenamefont {Casali}, \citenamefont {Geyer}, \citenamefont {Mason}, \citenamefont {Monteiro},\ and\ \citenamefont {Roehrig}}]{Casali:2015vta}%
  \BibitemOpen
  \bibfield  {author} {\bibinfo {author} {\bibfnamefont {E.}~\bibnamefont {Casali}}, \bibinfo {author} {\bibfnamefont {Y.}~\bibnamefont {Geyer}}, \bibinfo {author} {\bibfnamefont {L.}~\bibnamefont {Mason}}, \bibinfo {author} {\bibfnamefont {R.}~\bibnamefont {Monteiro}}, \ and\ \bibinfo {author} {\bibfnamefont {K.~A.}\ \bibnamefont {Roehrig}},\ }\href {\doibase 10.1007/JHEP11(2015)038} {\bibfield  {journal} {\bibinfo  {journal} {JHEP}\ }\textbf {\bibinfo {volume} {11}},\ \bibinfo {pages} {038} (\bibinfo {year} {2015})},\ \Eprint {http://arxiv.org/abs/1506.08771} {arXiv:1506.08771} \BibitemShut {NoStop}%
\bibitem [{\citenamefont {Bern}\ \emph {et~al.}(2008)\citenamefont {Bern}, \citenamefont {Carrasco},\ and\ \citenamefont {Johansson}}]{Bern:2008qj}%
  \BibitemOpen
  \bibfield  {author} {\bibinfo {author} {\bibfnamefont {Z.}~\bibnamefont {Bern}}, \bibinfo {author} {\bibfnamefont {J.~J.~M.}\ \bibnamefont {Carrasco}}, \ and\ \bibinfo {author} {\bibfnamefont {H.}~\bibnamefont {Johansson}},\ }\href {\doibase 10.1103/PhysRevD.78.085011} {\bibfield  {journal} {\bibinfo  {journal} {Phys. Rev. D}\ }\textbf {\bibinfo {volume} {78}},\ \bibinfo {pages} {085011} (\bibinfo {year} {2008})},\ \Eprint {http://arxiv.org/abs/0805.3993} {arXiv:0805.3993} \BibitemShut {NoStop}%
\bibitem [{\citenamefont {Bogers}\ and\ \citenamefont {Brauner}(2018)}]{Bogers:2018zeg}%
  \BibitemOpen
  \bibfield  {author} {\bibinfo {author} {\bibfnamefont {M.~P.}\ \bibnamefont {Bogers}}\ and\ \bibinfo {author} {\bibfnamefont {T.}~\bibnamefont {Brauner}},\ }\href {\doibase 10.1007/JHEP05(2018)076} {\bibfield  {journal} {\bibinfo  {journal} {JHEP}\ }\textbf {\bibinfo {volume} {05}},\ \bibinfo {pages} {076} (\bibinfo {year} {2018})},\ \Eprint {http://arxiv.org/abs/1803.05359} {arXiv:1803.05359 [hep-th]} \BibitemShut {NoStop}%
\bibitem [{\citenamefont {Roest}\ \emph {et~al.}(2019)\citenamefont {Roest}, \citenamefont {Stefanyszyn},\ and\ \citenamefont {Werkman}}]{Roest:2019oiw}%
  \BibitemOpen
  \bibfield  {author} {\bibinfo {author} {\bibfnamefont {D.}~\bibnamefont {Roest}}, \bibinfo {author} {\bibfnamefont {D.}~\bibnamefont {Stefanyszyn}}, \ and\ \bibinfo {author} {\bibfnamefont {P.}~\bibnamefont {Werkman}},\ }\href {\doibase 10.1007/JHEP08(2019)081} {\bibfield  {journal} {\bibinfo  {journal} {JHEP}\ }\textbf {\bibinfo {volume} {08}},\ \bibinfo {pages} {081} (\bibinfo {year} {2019})},\ \Eprint {http://arxiv.org/abs/1903.08222} {arXiv:1903.08222} \BibitemShut {NoStop}%
\bibitem [{\citenamefont {Kampf}\ \emph {et~al.}(2013{\natexlab{b}})\citenamefont {Kampf}, \citenamefont {Novotny},\ and\ \citenamefont {Trnka}}]{Kampf:2012fn}%
  \BibitemOpen
  \bibfield  {author} {\bibinfo {author} {\bibfnamefont {K.}~\bibnamefont {Kampf}}, \bibinfo {author} {\bibfnamefont {J.}~\bibnamefont {Novotny}}, \ and\ \bibinfo {author} {\bibfnamefont {J.}~\bibnamefont {Trnka}},\ }\href {\doibase 10.1103/PhysRevD.87.081701} {\bibfield  {journal} {\bibinfo  {journal} {Phys. Rev. D}\ }\textbf {\bibinfo {volume} {87}},\ \bibinfo {pages} {081701} (\bibinfo {year} {2013}{\natexlab{b}})},\ \Eprint {http://arxiv.org/abs/1212.5224} {arXiv:1212.5224} \BibitemShut {NoStop}%
\bibitem [{\citenamefont {Hinterbichler}\ and\ \citenamefont {Joyce}(2015)}]{Hinterbichler:2015pqa}%
  \BibitemOpen
  \bibfield  {author} {\bibinfo {author} {\bibfnamefont {K.}~\bibnamefont {Hinterbichler}}\ and\ \bibinfo {author} {\bibfnamefont {A.}~\bibnamefont {Joyce}},\ }\href {\doibase 10.1103/PhysRevD.92.023503} {\bibfield  {journal} {\bibinfo  {journal} {Phys. Rev. D}\ }\textbf {\bibinfo {volume} {92}},\ \bibinfo {pages} {023503} (\bibinfo {year} {2015})},\ \Eprint {http://arxiv.org/abs/1501.07600} {arXiv:1501.07600} \BibitemShut {NoStop}%
\bibitem [{\citenamefont {Arkani-Hamed}\ \emph {et~al.}(2023{\natexlab{a}})\citenamefont {Arkani-Hamed}, \citenamefont {Cao}, \citenamefont {Dong}, \citenamefont {Figueiredo},\ and\ \citenamefont {He}}]{Arkani-Hamed:2023swr}%
  \BibitemOpen
  \bibfield  {author} {\bibinfo {author} {\bibfnamefont {N.}~\bibnamefont {Arkani-Hamed}}, \bibinfo {author} {\bibfnamefont {Q.}~\bibnamefont {Cao}}, \bibinfo {author} {\bibfnamefont {J.}~\bibnamefont {Dong}}, \bibinfo {author} {\bibfnamefont {C.}~\bibnamefont {Figueiredo}}, \ and\ \bibinfo {author} {\bibfnamefont {S.}~\bibnamefont {He}},\ }\href@noop {} {\  (\bibinfo {year} {2023}{\natexlab{a}})},\ \Eprint {http://arxiv.org/abs/2312.16282} {arXiv:2312.16282} \BibitemShut {NoStop}%
\bibitem [{\citenamefont {Arkani-Hamed}\ \emph {et~al.}(2024{\natexlab{a}})\citenamefont {Arkani-Hamed}, \citenamefont {Cao}, \citenamefont {Dong}, \citenamefont {Figueiredo},\ and\ \citenamefont {He}}]{Arkani-Hamed:2024nhp}%
  \BibitemOpen
  \bibfield  {author} {\bibinfo {author} {\bibfnamefont {N.}~\bibnamefont {Arkani-Hamed}}, \bibinfo {author} {\bibfnamefont {Q.}~\bibnamefont {Cao}}, \bibinfo {author} {\bibfnamefont {J.}~\bibnamefont {Dong}}, \bibinfo {author} {\bibfnamefont {C.}~\bibnamefont {Figueiredo}}, \ and\ \bibinfo {author} {\bibfnamefont {S.}~\bibnamefont {He}},\ }\href@noop {} {\  (\bibinfo {year} {2024}{\natexlab{a}})},\ \Eprint {http://arxiv.org/abs/2401.05483} {arXiv:2401.05483 [hep-th]} \BibitemShut {NoStop}%
\bibitem [{\citenamefont {Arkani-Hamed}\ and\ \citenamefont {Figueiredo}(2024)}]{Arkani-Hamed:2024yvu}%
  \BibitemOpen
  \bibfield  {author} {\bibinfo {author} {\bibfnamefont {N.}~\bibnamefont {Arkani-Hamed}}\ and\ \bibinfo {author} {\bibfnamefont {C.}~\bibnamefont {Figueiredo}},\ }\href@noop {} {\  (\bibinfo {year} {2024})},\ \Eprint {http://arxiv.org/abs/2403.04826} {arXiv:2403.04826 [hep-th]} \BibitemShut {NoStop}%
\bibitem [{\citenamefont {Dixon}\ \emph {et~al.}(1999)\citenamefont {Dixon}, \citenamefont {Kunszt},\ and\ \citenamefont {Signer}}]{Dixon:1999di}%
  \BibitemOpen
  \bibfield  {author} {\bibinfo {author} {\bibfnamefont {L.~J.}\ \bibnamefont {Dixon}}, \bibinfo {author} {\bibfnamefont {Z.}~\bibnamefont {Kunszt}}, \ and\ \bibinfo {author} {\bibfnamefont {A.}~\bibnamefont {Signer}},\ }\href {\doibase 10.1103/PhysRevD.60.114037} {\bibfield  {journal} {\bibinfo  {journal} {Phys. Rev. D}\ }\textbf {\bibinfo {volume} {60}},\ \bibinfo {pages} {114037} (\bibinfo {year} {1999})},\ \Eprint {http://arxiv.org/abs/hep-ph/9907305} {arXiv:hep-ph/9907305} \BibitemShut {NoStop}%
\bibitem [{\citenamefont {Arkani-Hamed}\ \emph {et~al.}(2022)\citenamefont {Arkani-Hamed}, \citenamefont {He}, \citenamefont {Salvatori},\ and\ \citenamefont {Thomas}}]{Arkani-Hamed:2019vag}%
  \BibitemOpen
  \bibfield  {author} {\bibinfo {author} {\bibfnamefont {N.}~\bibnamefont {Arkani-Hamed}}, \bibinfo {author} {\bibfnamefont {S.}~\bibnamefont {He}}, \bibinfo {author} {\bibfnamefont {G.}~\bibnamefont {Salvatori}}, \ and\ \bibinfo {author} {\bibfnamefont {H.}~\bibnamefont {Thomas}},\ }\href {\doibase 10.1007/JHEP11(2022)049} {\bibfield  {journal} {\bibinfo  {journal} {JHEP}\ }\textbf {\bibinfo {volume} {11}},\ \bibinfo {pages} {049} (\bibinfo {year} {2022})},\ \Eprint {http://arxiv.org/abs/1912.12948} {arXiv:1912.12948 [hep-th]} \BibitemShut {NoStop}%
\bibitem [{\citenamefont {Arkani-Hamed}\ \emph {et~al.}(2023{\natexlab{b}})\citenamefont {Arkani-Hamed}, \citenamefont {Frost}, \citenamefont {Salvatori}, \citenamefont {Plamondon},\ and\ \citenamefont {Thomas}}]{Arkani-Hamed:2023lbd}%
  \BibitemOpen
  \bibfield  {author} {\bibinfo {author} {\bibfnamefont {N.}~\bibnamefont {Arkani-Hamed}}, \bibinfo {author} {\bibfnamefont {H.}~\bibnamefont {Frost}}, \bibinfo {author} {\bibfnamefont {G.}~\bibnamefont {Salvatori}}, \bibinfo {author} {\bibfnamefont {P.-G.}\ \bibnamefont {Plamondon}}, \ and\ \bibinfo {author} {\bibfnamefont {H.}~\bibnamefont {Thomas}},\ }\href@noop {} {\  (\bibinfo {year} {2023}{\natexlab{b}})},\ \Eprint {http://arxiv.org/abs/2309.15913} {arXiv:2309.15913} \BibitemShut {NoStop}%
\bibitem [{\citenamefont {Arkani-Hamed}\ \emph {et~al.}(2023{\natexlab{c}})\citenamefont {Arkani-Hamed}, \citenamefont {Frost}, \citenamefont {Salvatori}, \citenamefont {Plamondon},\ and\ \citenamefont {Thomas}}]{Arkani-Hamed:2023mvg}%
  \BibitemOpen
  \bibfield  {author} {\bibinfo {author} {\bibfnamefont {N.}~\bibnamefont {Arkani-Hamed}}, \bibinfo {author} {\bibfnamefont {H.}~\bibnamefont {Frost}}, \bibinfo {author} {\bibfnamefont {G.}~\bibnamefont {Salvatori}}, \bibinfo {author} {\bibfnamefont {P.-G.}\ \bibnamefont {Plamondon}}, \ and\ \bibinfo {author} {\bibfnamefont {H.}~\bibnamefont {Thomas}},\ }\href@noop {} {\  (\bibinfo {year} {2023}{\natexlab{c}})},\ \Eprint {http://arxiv.org/abs/2311.09284} {arXiv:2311.09284} \BibitemShut {NoStop}%
\bibitem [{\citenamefont {Arkani-Hamed}\ \emph {et~al.}(2024{\natexlab{b}})\citenamefont {Arkani-Hamed}, \citenamefont {Figueiredo}, \citenamefont {Frost},\ and\ \citenamefont {Salvatori}}]{Arkani-Hamed:2024vna}%
  \BibitemOpen
  \bibfield  {author} {\bibinfo {author} {\bibfnamefont {N.}~\bibnamefont {Arkani-Hamed}}, \bibinfo {author} {\bibfnamefont {C.}~\bibnamefont {Figueiredo}}, \bibinfo {author} {\bibfnamefont {H.}~\bibnamefont {Frost}}, \ and\ \bibinfo {author} {\bibfnamefont {G.}~\bibnamefont {Salvatori}},\ }\href@noop {} {\  (\bibinfo {year} {2024}{\natexlab{b}})},\ \Eprint {http://arxiv.org/abs/2402.06719} {arXiv:2402.06719 [hep-th]} \BibitemShut {NoStop}%
\bibitem [{\citenamefont {Cao}\ \emph {et~al.}(2024)\citenamefont {Cao}, \citenamefont {Dong}, \citenamefont {He},\ and\ \citenamefont {Shi}}]{Cao:2024uni}%
  \BibitemOpen
  \bibfield  {author} {\bibinfo {author} {\bibfnamefont {Q.}~\bibnamefont {Cao}}, \bibinfo {author} {\bibfnamefont {J.}~\bibnamefont {Dong}}, \bibinfo {author} {\bibfnamefont {S.}~\bibnamefont {He}}, \ and\ \bibinfo {author} {\bibfnamefont {C.}~\bibnamefont {Shi}},\ }\href@noop {} {\  (\bibinfo {year} {2024})},\ \Eprint {http://arxiv.org/abs/2403.08855} {arXiv:2403.08855 [hep-th]} \BibitemShut {NoStop}%
\bibitem [{\citenamefont {Bern}\ \emph {et~al.}(2019)\citenamefont {Bern}, \citenamefont {Carrasco}, \citenamefont {Chiodaroli}, \citenamefont {Johansson},\ and\ \citenamefont {Roiban}}]{Bern:2019prr}%
  \BibitemOpen
  \bibfield  {author} {\bibinfo {author} {\bibfnamefont {Z.}~\bibnamefont {Bern}}, \bibinfo {author} {\bibfnamefont {J.~J.}\ \bibnamefont {Carrasco}}, \bibinfo {author} {\bibfnamefont {M.}~\bibnamefont {Chiodaroli}}, \bibinfo {author} {\bibfnamefont {H.}~\bibnamefont {Johansson}}, \ and\ \bibinfo {author} {\bibfnamefont {R.}~\bibnamefont {Roiban}},\ }\href@noop {} {\  (\bibinfo {year} {2019})},\ \Eprint {http://arxiv.org/abs/1909.01358} {arXiv:1909.01358} \BibitemShut {NoStop}%
\bibitem [{\citenamefont {Low}\ and\ \citenamefont {Yin}(2019)}]{Low:2019ynd}%
  \BibitemOpen
  \bibfield  {author} {\bibinfo {author} {\bibfnamefont {I.}~\bibnamefont {Low}}\ and\ \bibinfo {author} {\bibfnamefont {Z.}~\bibnamefont {Yin}},\ }\href {\doibase 10.1007/JHEP11(2019)078} {\bibfield  {journal} {\bibinfo  {journal} {JHEP}\ }\textbf {\bibinfo {volume} {11}},\ \bibinfo {pages} {078} (\bibinfo {year} {2019})},\ \Eprint {http://arxiv.org/abs/1904.12859} {arXiv:1904.12859} \BibitemShut {NoStop}%
\bibitem [{\citenamefont {Brown}\ \emph {et~al.}(2023)\citenamefont {Brown}, \citenamefont {Kampf}, \citenamefont {Oktem}, \citenamefont {Paranjape},\ and\ \citenamefont {Trnka}}]{Brown:2023srz}%
  \BibitemOpen
  \bibfield  {author} {\bibinfo {author} {\bibfnamefont {T.~V.}\ \bibnamefont {Brown}}, \bibinfo {author} {\bibfnamefont {K.}~\bibnamefont {Kampf}}, \bibinfo {author} {\bibfnamefont {U.}~\bibnamefont {Oktem}}, \bibinfo {author} {\bibfnamefont {S.}~\bibnamefont {Paranjape}}, \ and\ \bibinfo {author} {\bibfnamefont {J.}~\bibnamefont {Trnka}},\ }\href {\doibase 10.1103/PhysRevD.108.105008} {\bibfield  {journal} {\bibinfo  {journal} {Phys. Rev. D}\ }\textbf {\bibinfo {volume} {108}},\ \bibinfo {pages} {105008} (\bibinfo {year} {2023})},\ \Eprint {http://arxiv.org/abs/2305.05688} {arXiv:2305.05688} \BibitemShut {NoStop}%
\bibitem [{\citenamefont {Kampf}\ \emph {et~al.}(2021)\citenamefont {Kampf}, \citenamefont {Novotny}, \citenamefont {Preucil},\ and\ \citenamefont {Trnka}}]{Kampf:2021bet}%
  \BibitemOpen
  \bibfield  {author} {\bibinfo {author} {\bibfnamefont {K.}~\bibnamefont {Kampf}}, \bibinfo {author} {\bibfnamefont {J.}~\bibnamefont {Novotny}}, \bibinfo {author} {\bibfnamefont {F.}~\bibnamefont {Preucil}}, \ and\ \bibinfo {author} {\bibfnamefont {J.}~\bibnamefont {Trnka}},\ }\href {\doibase 10.1007/JHEP07(2021)153} {\bibfield  {journal} {\bibinfo  {journal} {JHEP}\ }\textbf {\bibinfo {volume} {07}},\ \bibinfo {pages} {153} (\bibinfo {year} {2021})},\ \Eprint {http://arxiv.org/abs/2104.10693} {arXiv:2104.10693} \BibitemShut {NoStop}%
\bibitem [{\citenamefont {Kampf}(2021)}]{Kampf:2021jvf}%
  \BibitemOpen
  \bibfield  {author} {\bibinfo {author} {\bibfnamefont {K.}~\bibnamefont {Kampf}},\ }\href {\doibase 10.1007/JHEP12(2021)140} {\bibfield  {journal} {\bibinfo  {journal} {JHEP}\ }\textbf {\bibinfo {volume} {12}},\ \bibinfo {pages} {140} (\bibinfo {year} {2021})},\ \Eprint {http://arxiv.org/abs/2109.11574} {arXiv:2109.11574} \BibitemShut {NoStop}%
\bibitem [{\citenamefont {Bijnens}\ \emph {et~al.}(2019)\citenamefont {Bijnens}, \citenamefont {Kampf},\ and\ \citenamefont {Sj\"o}}]{Bijnens:2019eze}%
  \BibitemOpen
  \bibfield  {author} {\bibinfo {author} {\bibfnamefont {J.}~\bibnamefont {Bijnens}}, \bibinfo {author} {\bibfnamefont {K.}~\bibnamefont {Kampf}}, \ and\ \bibinfo {author} {\bibfnamefont {M.}~\bibnamefont {Sj\"o}},\ }\href {\doibase 10.1007/JHEP11(2019)074} {\bibfield  {journal} {\bibinfo  {journal} {JHEP}\ }\textbf {\bibinfo {volume} {11}},\ \bibinfo {pages} {074} (\bibinfo {year} {2019})},\ \bibinfo {note} {[Erratum: JHEP 03, 066 (2021)]},\ \Eprint {http://arxiv.org/abs/1909.13684} {arXiv:1909.13684} \BibitemShut {NoStop}%
\bibitem [{\citenamefont {Cachazo}\ \emph {et~al.}(2016)\citenamefont {Cachazo}, \citenamefont {Cha},\ and\ \citenamefont {Mizera}}]{Cachazo:2016njl}%
  \BibitemOpen
  \bibfield  {author} {\bibinfo {author} {\bibfnamefont {F.}~\bibnamefont {Cachazo}}, \bibinfo {author} {\bibfnamefont {P.}~\bibnamefont {Cha}}, \ and\ \bibinfo {author} {\bibfnamefont {S.}~\bibnamefont {Mizera}},\ }\href {\doibase 10.1007/JHEP06(2016)170} {\bibfield  {journal} {\bibinfo  {journal} {JHEP}\ }\textbf {\bibinfo {volume} {06}},\ \bibinfo {pages} {170} (\bibinfo {year} {2016})},\ \Eprint {http://arxiv.org/abs/1604.03893} {arXiv:1604.03893} \BibitemShut {NoStop}%
\bibitem [{\citenamefont {Kawai}\ \emph {et~al.}(1986)\citenamefont {Kawai}, \citenamefont {Lewellen},\ and\ \citenamefont {Tye}}]{Kawai:1985xq}%
  \BibitemOpen
  \bibfield  {author} {\bibinfo {author} {\bibfnamefont {H.}~\bibnamefont {Kawai}}, \bibinfo {author} {\bibfnamefont {D.~C.}\ \bibnamefont {Lewellen}}, \ and\ \bibinfo {author} {\bibfnamefont {S.~H.~H.}\ \bibnamefont {Tye}},\ }\href {\doibase 10.1016/0550-3213(86)90362-7} {\bibfield  {journal} {\bibinfo  {journal} {Nucl. Phys. B}\ }\textbf {\bibinfo {volume} {269}},\ \bibinfo {pages} {1} (\bibinfo {year} {1986})}\BibitemShut {NoStop}%
\bibitem [{\citenamefont {Bjerrum-Bohr}\ \emph {et~al.}(2011)\citenamefont {Bjerrum-Bohr}, \citenamefont {Damgaard}, \citenamefont {Sondergaard},\ and\ \citenamefont {Vanhove}}]{Bjerrum-Bohr:2010pnr}%
  \BibitemOpen
  \bibfield  {author} {\bibinfo {author} {\bibfnamefont {N.~E.~J.}\ \bibnamefont {Bjerrum-Bohr}}, \bibinfo {author} {\bibfnamefont {P.~H.}\ \bibnamefont {Damgaard}}, \bibinfo {author} {\bibfnamefont {T.}~\bibnamefont {Sondergaard}}, \ and\ \bibinfo {author} {\bibfnamefont {P.}~\bibnamefont {Vanhove}},\ }\href {\doibase 10.1007/JHEP01(2011)001} {\bibfield  {journal} {\bibinfo  {journal} {JHEP}\ }\textbf {\bibinfo {volume} {01}},\ \bibinfo {pages} {001} (\bibinfo {year} {2011})},\ \Eprint {http://arxiv.org/abs/1010.3933} {arXiv:1010.3933} \BibitemShut {NoStop}%
\bibitem [{\citenamefont {Bartsch}\ \emph {et~al.}()\citenamefont {Bartsch}, \citenamefont {Jones}, \citenamefont {Kampf}, \citenamefont {Oktem}, \citenamefont {Paranjape},\ and\ \citenamefont {Trnka}}]{inprogress:Bartsch}%
  \BibitemOpen
  \bibfield  {author} {\bibinfo {author} {\bibfnamefont {C.}~\bibnamefont {Bartsch}}, \bibinfo {author} {\bibfnamefont {C.}~\bibnamefont {Jones}}, \bibinfo {author} {\bibfnamefont {K.}~\bibnamefont {Kampf}}, \bibinfo {author} {\bibfnamefont {U.}~\bibnamefont {Oktem}}, \bibinfo {author} {\bibfnamefont {S.}~\bibnamefont {Paranjape}}, \ and\ \bibinfo {author} {\bibfnamefont {J.}~\bibnamefont {Trnka}},\ }\href@noop {} {\ }\Eprint {http://arxiv.org/abs/(in progress)} {(in progress)} \BibitemShut {NoStop}%
\bibitem [{\citenamefont {Carrillo~Gonz\'alez}\ \emph {et~al.}(2020)\citenamefont {Carrillo~Gonz\'alez}, \citenamefont {Penco},\ and\ \citenamefont {Trodden}}]{CarrilloGonzalez:2019fzc}%
  \BibitemOpen
  \bibfield  {author} {\bibinfo {author} {\bibfnamefont {M.}~\bibnamefont {Carrillo~Gonz\'alez}}, \bibinfo {author} {\bibfnamefont {R.}~\bibnamefont {Penco}}, \ and\ \bibinfo {author} {\bibfnamefont {M.}~\bibnamefont {Trodden}},\ }\href {\doibase 10.1103/PhysRevD.102.105011} {\bibfield  {journal} {\bibinfo  {journal} {Phys. Rev. D}\ }\textbf {\bibinfo {volume} {102}},\ \bibinfo {pages} {105011} (\bibinfo {year} {2020})},\ \Eprint {http://arxiv.org/abs/1908.07531} {arXiv:1908.07531} \BibitemShut {NoStop}%
\bibitem [{\citenamefont {Chi}\ \emph {et~al.}(2022)\citenamefont {Chi}, \citenamefont {Elvang}, \citenamefont {Herderschee}, \citenamefont {Jones},\ and\ \citenamefont {Paranjape}}]{Chi:2021mio}%
  \BibitemOpen
  \bibfield  {author} {\bibinfo {author} {\bibfnamefont {H.-H.}\ \bibnamefont {Chi}}, \bibinfo {author} {\bibfnamefont {H.}~\bibnamefont {Elvang}}, \bibinfo {author} {\bibfnamefont {A.}~\bibnamefont {Herderschee}}, \bibinfo {author} {\bibfnamefont {C.~R.~T.}\ \bibnamefont {Jones}}, \ and\ \bibinfo {author} {\bibfnamefont {S.}~\bibnamefont {Paranjape}},\ }\href {\doibase 10.1007/JHEP03(2022)077} {\bibfield  {journal} {\bibinfo  {journal} {JHEP}\ }\textbf {\bibinfo {volume} {03}},\ \bibinfo {pages} {077} (\bibinfo {year} {2022})},\ \Eprint {http://arxiv.org/abs/2106.12600} {arXiv:2106.12600} \BibitemShut {NoStop}%
\bibitem [{\citenamefont {Chen}\ \emph {et~al.}(2023)\citenamefont {Chen}, \citenamefont {Elvang},\ and\ \citenamefont {Herderschee}}]{Chen:2023dcx}%
  \BibitemOpen
  \bibfield  {author} {\bibinfo {author} {\bibfnamefont {A.~S.-K.}\ \bibnamefont {Chen}}, \bibinfo {author} {\bibfnamefont {H.}~\bibnamefont {Elvang}}, \ and\ \bibinfo {author} {\bibfnamefont {A.}~\bibnamefont {Herderschee}},\ }\href@noop {} {\  (\bibinfo {year} {2023})},\ \Eprint {http://arxiv.org/abs/2302.04895} {arXiv:2302.04895} \BibitemShut {NoStop}%
\bibitem [{\citenamefont {Chen}\ \emph {et~al.}(2022)\citenamefont {Chen}, \citenamefont {Elvang},\ and\ \citenamefont {Herderschee}}]{Chen:2022shl}%
  \BibitemOpen
  \bibfield  {author} {\bibinfo {author} {\bibfnamefont {A.~S.-K.}\ \bibnamefont {Chen}}, \bibinfo {author} {\bibfnamefont {H.}~\bibnamefont {Elvang}}, \ and\ \bibinfo {author} {\bibfnamefont {A.}~\bibnamefont {Herderschee}},\ }\href@noop {} {\  (\bibinfo {year} {2022})},\ \Eprint {http://arxiv.org/abs/2212.13998} {arXiv:2212.13998} \BibitemShut {NoStop}%
\bibitem [{\citenamefont {Carrasco}\ \emph {et~al.}(2020)\citenamefont {Carrasco}, \citenamefont {Rodina}, \citenamefont {Yin},\ and\ \citenamefont {Zekioglu}}]{Carrasco:2019yyn}%
  \BibitemOpen
  \bibfield  {author} {\bibinfo {author} {\bibfnamefont {J.~J.~M.}\ \bibnamefont {Carrasco}}, \bibinfo {author} {\bibfnamefont {L.}~\bibnamefont {Rodina}}, \bibinfo {author} {\bibfnamefont {Z.}~\bibnamefont {Yin}}, \ and\ \bibinfo {author} {\bibfnamefont {S.}~\bibnamefont {Zekioglu}},\ }\href {\doibase 10.1103/PhysRevLett.125.251602} {\bibfield  {journal} {\bibinfo  {journal} {Phys. Rev. Lett.}\ }\textbf {\bibinfo {volume} {125}},\ \bibinfo {pages} {251602} (\bibinfo {year} {2020})},\ \Eprint {http://arxiv.org/abs/1910.12850} {arXiv:1910.12850} \BibitemShut {NoStop}%
\bibitem [{\citenamefont {Carrasco}\ \emph {et~al.}(2021)\citenamefont {Carrasco}, \citenamefont {Rodina},\ and\ \citenamefont {Zekioglu}}]{Carrasco:2021ptp}%
  \BibitemOpen
  \bibfield  {author} {\bibinfo {author} {\bibfnamefont {J.~J.~M.}\ \bibnamefont {Carrasco}}, \bibinfo {author} {\bibfnamefont {L.}~\bibnamefont {Rodina}}, \ and\ \bibinfo {author} {\bibfnamefont {S.}~\bibnamefont {Zekioglu}},\ }\href {\doibase 10.1007/JHEP06(2021)169} {\bibfield  {journal} {\bibinfo  {journal} {JHEP}\ }\textbf {\bibinfo {volume} {06}},\ \bibinfo {pages} {169} (\bibinfo {year} {2021})},\ \Eprint {http://arxiv.org/abs/2104.08370} {arXiv:2104.08370} \BibitemShut {NoStop}%
\bibitem [{\citenamefont {Arkani-Hamed}\ \emph {et~al.}(2024{\natexlab{c}})\citenamefont {Arkani-Hamed}, \citenamefont {Cao}, \citenamefont {Dong}, \citenamefont {Figueiredo},\ and\ \citenamefont {He}}]{nima2}%
  \BibitemOpen
  \bibfield  {author} {\bibinfo {author} {\bibfnamefont {N.}~\bibnamefont {Arkani-Hamed}}, \bibinfo {author} {\bibfnamefont {Q.}~\bibnamefont {Cao}}, \bibinfo {author} {\bibfnamefont {J.}~\bibnamefont {Dong}}, \bibinfo {author} {\bibfnamefont {C.}~\bibnamefont {Figueiredo}}, \ and\ \bibinfo {author} {\bibfnamefont {S.}~\bibnamefont {He}},\ }\href@noop {} {\  (\bibinfo {year} {2024}{\natexlab{c}})},\ \Eprint {http://arxiv.org/abs/2401.05483} {arXiv:2401.05483} \BibitemShut {NoStop}%
\bibitem [{\citenamefont {Arkani-Hamed}\ \emph {et~al.}(2018{\natexlab{a}})\citenamefont {Arkani-Hamed}, \citenamefont {Bai}, \citenamefont {He},\ and\ \citenamefont {Yan}}]{Arkani-Hamed:2017mur}%
  \BibitemOpen
  \bibfield  {author} {\bibinfo {author} {\bibfnamefont {N.}~\bibnamefont {Arkani-Hamed}}, \bibinfo {author} {\bibfnamefont {Y.}~\bibnamefont {Bai}}, \bibinfo {author} {\bibfnamefont {S.}~\bibnamefont {He}}, \ and\ \bibinfo {author} {\bibfnamefont {G.}~\bibnamefont {Yan}},\ }\href {\doibase 10.1007/JHEP05(2018)096} {\bibfield  {journal} {\bibinfo  {journal} {JHEP}\ }\textbf {\bibinfo {volume} {05}},\ \bibinfo {pages} {096} (\bibinfo {year} {2018}{\natexlab{a}})},\ \Eprint {http://arxiv.org/abs/1711.09102} {arXiv:1711.09102} \BibitemShut {NoStop}%
\bibitem [{\citenamefont {Arkani-Hamed}\ and\ \citenamefont {Trnka}(2014{\natexlab{a}})}]{Arkani-Hamed:2013jha}%
  \BibitemOpen
  \bibfield  {author} {\bibinfo {author} {\bibfnamefont {N.}~\bibnamefont {Arkani-Hamed}}\ and\ \bibinfo {author} {\bibfnamefont {J.}~\bibnamefont {Trnka}},\ }\href {\doibase 10.1007/JHEP10(2014)030} {\bibfield  {journal} {\bibinfo  {journal} {JHEP}\ }\textbf {\bibinfo {volume} {10}},\ \bibinfo {pages} {030} (\bibinfo {year} {2014}{\natexlab{a}})},\ \Eprint {http://arxiv.org/abs/1312.2007} {arXiv:1312.2007} \BibitemShut {NoStop}%
\bibitem [{\citenamefont {Arkani-Hamed}\ and\ \citenamefont {Trnka}(2014{\natexlab{b}})}]{Arkani-Hamed:2013kca}%
  \BibitemOpen
  \bibfield  {author} {\bibinfo {author} {\bibfnamefont {N.}~\bibnamefont {Arkani-Hamed}}\ and\ \bibinfo {author} {\bibfnamefont {J.}~\bibnamefont {Trnka}},\ }\href {\doibase 10.1007/JHEP12(2014)182} {\bibfield  {journal} {\bibinfo  {journal} {JHEP}\ }\textbf {\bibinfo {volume} {12}},\ \bibinfo {pages} {182} (\bibinfo {year} {2014}{\natexlab{b}})},\ \Eprint {http://arxiv.org/abs/1312.7878} {arXiv:1312.7878 [hep-th]} \BibitemShut {NoStop}%
\bibitem [{\citenamefont {Arkani-Hamed}\ \emph {et~al.}(2018{\natexlab{b}})\citenamefont {Arkani-Hamed}, \citenamefont {Thomas},\ and\ \citenamefont {Trnka}}]{Arkani-Hamed:2017vfh}%
  \BibitemOpen
  \bibfield  {author} {\bibinfo {author} {\bibfnamefont {N.}~\bibnamefont {Arkani-Hamed}}, \bibinfo {author} {\bibfnamefont {H.}~\bibnamefont {Thomas}}, \ and\ \bibinfo {author} {\bibfnamefont {J.}~\bibnamefont {Trnka}},\ }\href {\doibase 10.1007/JHEP01(2018)016} {\bibfield  {journal} {\bibinfo  {journal} {JHEP}\ }\textbf {\bibinfo {volume} {01}},\ \bibinfo {pages} {016} (\bibinfo {year} {2018}{\natexlab{b}})},\ \Eprint {http://arxiv.org/abs/1704.05069} {arXiv:1704.05069} \BibitemShut {NoStop}%
\bibitem [{\citenamefont {Damgaard}\ \emph {et~al.}(2019)\citenamefont {Damgaard}, \citenamefont {Ferro}, \citenamefont {Lukowski},\ and\ \citenamefont {Parisi}}]{Damgaard:2019ztj}%
  \BibitemOpen
  \bibfield  {author} {\bibinfo {author} {\bibfnamefont {D.}~\bibnamefont {Damgaard}}, \bibinfo {author} {\bibfnamefont {L.}~\bibnamefont {Ferro}}, \bibinfo {author} {\bibfnamefont {T.}~\bibnamefont {Lukowski}}, \ and\ \bibinfo {author} {\bibfnamefont {M.}~\bibnamefont {Parisi}},\ }\href {\doibase 10.1007/JHEP08(2019)042} {\bibfield  {journal} {\bibinfo  {journal} {JHEP}\ }\textbf {\bibinfo {volume} {08}},\ \bibinfo {pages} {042} (\bibinfo {year} {2019})},\ \Eprint {http://arxiv.org/abs/1905.04216} {arXiv:1905.04216} \BibitemShut {NoStop}%
\bibitem [{\citenamefont {He}\ \emph {et~al.}(2023)\citenamefont {He}, \citenamefont {Huang},\ and\ \citenamefont {Kuo}}]{He:2023rou}%
  \BibitemOpen
  \bibfield  {author} {\bibinfo {author} {\bibfnamefont {S.}~\bibnamefont {He}}, \bibinfo {author} {\bibfnamefont {Y.-t.}\ \bibnamefont {Huang}}, \ and\ \bibinfo {author} {\bibfnamefont {C.-K.}\ \bibnamefont {Kuo}},\ }\href {\doibase 10.1007/JHEP09(2023)165} {\bibfield  {journal} {\bibinfo  {journal} {JHEP}\ }\textbf {\bibinfo {volume} {09}},\ \bibinfo {pages} {165} (\bibinfo {year} {2023})},\ \Eprint {http://arxiv.org/abs/2306.00951} {arXiv:2306.00951 [hep-th]} \BibitemShut {NoStop}%
\bibitem [{\citenamefont {Arkani-Hamed}\ \emph {et~al.}(2018{\natexlab{c}})\citenamefont {Arkani-Hamed}, \citenamefont {Rodina},\ and\ \citenamefont {Trnka}}]{Arkani-Hamed:2016rak}%
  \BibitemOpen
  \bibfield  {author} {\bibinfo {author} {\bibfnamefont {N.}~\bibnamefont {Arkani-Hamed}}, \bibinfo {author} {\bibfnamefont {L.}~\bibnamefont {Rodina}}, \ and\ \bibinfo {author} {\bibfnamefont {J.}~\bibnamefont {Trnka}},\ }\href {\doibase 10.1103/PhysRevLett.120.231602} {\bibfield  {journal} {\bibinfo  {journal} {Phys. Rev. Lett.}\ }\textbf {\bibinfo {volume} {120}},\ \bibinfo {pages} {231602} (\bibinfo {year} {2018}{\natexlab{c}})},\ \Eprint {http://arxiv.org/abs/1612.02797} {arXiv:1612.02797} \BibitemShut {NoStop}%
\bibitem [{\citenamefont {Rodina}(2019{\natexlab{a}})}]{Rodina:2016jyz}%
  \BibitemOpen
  \bibfield  {author} {\bibinfo {author} {\bibfnamefont {L.}~\bibnamefont {Rodina}},\ }\href {\doibase 10.1007/JHEP09(2019)084} {\bibfield  {journal} {\bibinfo  {journal} {JHEP}\ }\textbf {\bibinfo {volume} {09}},\ \bibinfo {pages} {084} (\bibinfo {year} {2019}{\natexlab{a}})},\ \Eprint {http://arxiv.org/abs/1612.06342} {arXiv:1612.06342} \BibitemShut {NoStop}%
\bibitem [{\citenamefont {Rodina}(2019{\natexlab{b}})}]{Rodina:2016mbk}%
  \BibitemOpen
  \bibfield  {author} {\bibinfo {author} {\bibfnamefont {L.}~\bibnamefont {Rodina}},\ }\href {\doibase 10.1007/JHEP09(2019)078} {\bibfield  {journal} {\bibinfo  {journal} {JHEP}\ }\textbf {\bibinfo {volume} {09}},\ \bibinfo {pages} {078} (\bibinfo {year} {2019}{\natexlab{b}})},\ \Eprint {http://arxiv.org/abs/1612.03885} {arXiv:1612.03885 [hep-th]} \BibitemShut {NoStop}%
\bibitem [{\citenamefont {Arkani-Hamed}\ \emph {et~al.}(2011)\citenamefont {Arkani-Hamed}, \citenamefont {Bourjaily}, \citenamefont {Cachazo}, \citenamefont {Caron-Huot},\ and\ \citenamefont {Trnka}}]{Arkani-Hamed:2010zjl}%
  \BibitemOpen
  \bibfield  {author} {\bibinfo {author} {\bibfnamefont {N.}~\bibnamefont {Arkani-Hamed}}, \bibinfo {author} {\bibfnamefont {J.~L.}\ \bibnamefont {Bourjaily}}, \bibinfo {author} {\bibfnamefont {F.}~\bibnamefont {Cachazo}}, \bibinfo {author} {\bibfnamefont {S.}~\bibnamefont {Caron-Huot}}, \ and\ \bibinfo {author} {\bibfnamefont {J.}~\bibnamefont {Trnka}},\ }\href {\doibase 10.1007/JHEP01(2011)041} {\bibfield  {journal} {\bibinfo  {journal} {JHEP}\ }\textbf {\bibinfo {volume} {01}},\ \bibinfo {pages} {041} (\bibinfo {year} {2011})},\ \Eprint {http://arxiv.org/abs/1008.2958} {arXiv:1008.2958} \BibitemShut {NoStop}%
\bibitem [{\citenamefont {Bartsch}\ \emph {et~al.}(2022)\citenamefont {Bartsch}, \citenamefont {Kampf},\ and\ \citenamefont {Trnka}}]{Bartsch:2022pyi}%
  \BibitemOpen
  \bibfield  {author} {\bibinfo {author} {\bibfnamefont {C.}~\bibnamefont {Bartsch}}, \bibinfo {author} {\bibfnamefont {K.}~\bibnamefont {Kampf}}, \ and\ \bibinfo {author} {\bibfnamefont {J.}~\bibnamefont {Trnka}},\ }\href {\doibase 10.1103/PhysRevD.106.076008} {\bibfield  {journal} {\bibinfo  {journal} {Phys. Rev. D}\ }\textbf {\bibinfo {volume} {106}},\ \bibinfo {pages} {076008} (\bibinfo {year} {2022})},\ \Eprint {http://arxiv.org/abs/2206.04694} {arXiv:2206.04694} \BibitemShut {NoStop}%
\bibitem [{\citenamefont {Bartsch}\ \emph {et~al.}(2024)\citenamefont {Bartsch}, \citenamefont {Kampf}, \citenamefont {Novotny},\ and\ \citenamefont {Trnka}}]{Bartsch:2024ofb}%
  \BibitemOpen
  \bibfield  {author} {\bibinfo {author} {\bibfnamefont {C.}~\bibnamefont {Bartsch}}, \bibinfo {author} {\bibfnamefont {K.}~\bibnamefont {Kampf}}, \bibinfo {author} {\bibfnamefont {J.}~\bibnamefont {Novotny}}, \ and\ \bibinfo {author} {\bibfnamefont {J.}~\bibnamefont {Trnka}},\ }\href@noop {} {\  (\bibinfo {year} {2024})},\ \Eprint {http://arxiv.org/abs/2401.04731} {arXiv:2401.04731 [hep-th]} \BibitemShut {NoStop}%
\bibitem [{\citenamefont {Luo}\ and\ \citenamefont {Wen}(2016)}]{Luo:2015tat}%
  \BibitemOpen
  \bibfield  {author} {\bibinfo {author} {\bibfnamefont {H.}~\bibnamefont {Luo}}\ and\ \bibinfo {author} {\bibfnamefont {C.}~\bibnamefont {Wen}},\ }\href {\doibase 10.1007/JHEP03(2016)088} {\bibfield  {journal} {\bibinfo  {journal} {JHEP}\ }\textbf {\bibinfo {volume} {03}},\ \bibinfo {pages} {088} (\bibinfo {year} {2016})},\ \Eprint {http://arxiv.org/abs/1512.06801} {arXiv:1512.06801} \BibitemShut {NoStop}%
\bibitem [{\citenamefont {Kampf}\ \emph {et~al.}(2020)\citenamefont {Kampf}, \citenamefont {Novotny}, \citenamefont {Shifman},\ and\ \citenamefont {Trnka}}]{Kampf:2019mcd}%
  \BibitemOpen
  \bibfield  {author} {\bibinfo {author} {\bibfnamefont {K.}~\bibnamefont {Kampf}}, \bibinfo {author} {\bibfnamefont {J.}~\bibnamefont {Novotny}}, \bibinfo {author} {\bibfnamefont {M.}~\bibnamefont {Shifman}}, \ and\ \bibinfo {author} {\bibfnamefont {J.}~\bibnamefont {Trnka}},\ }\href {\doibase 10.1103/PhysRevLett.124.111601} {\bibfield  {journal} {\bibinfo  {journal} {Phys. Rev. Lett.}\ }\textbf {\bibinfo {volume} {124}},\ \bibinfo {pages} {111601} (\bibinfo {year} {2020})},\ \Eprint {http://arxiv.org/abs/1910.04766} {arXiv:1910.04766} \BibitemShut {NoStop}%
\end{thebibliography}%
\bibliographystyle{apsrev4-1}

\end{document}